\documentclass[showpacs,amsmath,amssymb,superscriptaddress,twocolumn,prd]{revtex4}
\input{tcilatex}
\usepackage{graphicx,amsmath,color}
\usepackage{graphicx}
\usepackage{psfrag}
\newcommand{\ve}{\varepsilon}
\newcommand{\vp}{\varphi}

\begin{document}
\title{Curvature corrections and topology change transition\\ in brane-black hole systems: A perturbative approach}

\author{Viktor G. Czinner}
\email{czinner@rmki.kfki.hu}
\affiliation{Theoretical Physics Center for Science Facilities, IHEP CAS,\\ 
P.O.~Box 918-4, Beijing 100049, China}
\affiliation{KFKI Research Institute for Particle and Nuclear Physics, Department of Theoretical Physics,\\ Budapest 114, P.O.~Box 49, H-1525, Hungary}
\author{Antonino Flachi}
\email{flachi@yukawa.kyoto-u.ac.jp}
\affiliation{Yukawa Institute for Theoretical Physics, Kyoto University, Kyoto 606-8502, Japan}

\date{\today}

\begin{abstract}
We consider curvature corrections to static, axisymmetric Dirac-Nambu-Goto membranes 
embedded into a spherically symmetric black hole spacetime with arbitrary number of dimensions. Since the next to leading order corrections in the effective brane action are quadratic in the brane thickness $\ell$, we adopt a linear perturbation approach in $\ell^2$. The perturbations are general in the sense that they are not restricted to the Rindler zone nor to the near-critical solutions of the unperturbed system. As a result, an unexpected asymmetry in the perturbed system is found. In configurations, where the brane does not cross the black hole horizon, the perturbative approach does not lead to regular solutions if the number of the brane's spacetime dimensions $D>3$. This condition, however, does not hold for the horizon crossing solutions. Consequently we argue that the presented perturbative approach breaks down for subcritical type solutions near the axis of the system for $D>3$. Nevertheless, we can discuss topology-changing phase transitions in cases when $D=2$ or $3$, i.e.~when the brane is a 1-dimensional string or a 2-dimensional sheet, respectively. For the general case, a different, non-perturbative approach should be sought. Based on the energy properties of those branes that are quasi-statically evolved from the equatorial configuration, we illustrate the results of the phase transition in the case of a $D=3$ brane. It is found that small thickness perturbations do not modify the order of the transition, i.e.~it remains first order just as in the case of vanishing thickness.
\end{abstract}


\maketitle
\section{Introduction}
Higher dimensional black objects and branes are of importance and interest in several areas 
of present days physics. The classical black hole uniqueness theorems are known to fail in 
higher dimensions, and it turns out that a whole menagerie of black objects (black strings, rings, cigars, etc.) appear to exist. The study of new types of black objects became a very active research field recently, and among many other interesting aspects, the properties of possible transitions between the different types, or {\it phases}, is of special interest (see e.g. \cite{Kol1, Kol2} and references therein). For example, during the transition between a caged black hole and a black string phase, Kol \cite{Kol3} demonstrated that the Euclidean topology of the system changes. This type of transition is called {\it merger} transition, and Kol found a strong similarity in its properties with the Choptuik critical collapse phenomena \cite{Choptuik}. 

Recently, Frolov suggested a simple toy model with many features in common with merger and topology changing transitions \cite{Frolov}. The model consists of a bulk $N$-dimensional black hole and a test $D$-dimensional brane in it ($D\leq N-1$), called {\it brane-black hole} (BBH) system. The black hole is spherically symmetric, static and can be neutral or charged. The brane is infinitely thin, and it is described by the Dirac-Nambu-Goto \cite{Dirac, Nambu, Goto} action. It is also static, spherically symmetric and it is assumed to reach asymptotic infinity in the form of a $(D-1)$-dimensional plane. Due to the gravitational attraction of the black hole, the brane is deformed and there are two types of 
equilibrium configurations. The brane either crosses the black hole horizon, or it lies totally outside of the black hole (see FIG.~1). In between the two types of configurations there 
exists a critical solution that separates the two phases. Frolov studied the transition between the so called {\it subcritical phase} (when the brane does not intersect the black hole horizon) and {\it supercritical phase} (when the brane crosses the horizon), and found a close similarity both with the merger transition in a caged black hole - black string system and with the Choptuik critical collapse phenomena.

The AdS/CFT correspondence \cite{Maldacena} also provides motivation to study the above BBH system. In fact, according to the correspondence, 
at sufficiently high temperature, a small number of flavors ($N_f$) of fundamental matter in strongly coupled gauge theories with a large number of colors ($N_c\gg N_f$), may be described, in the holographic dual, by probe Dq-branes in the gravitational background of a black hole \cite{Witten, KR, KK}. In \cite{MMT1, MMT2}, using the tool of the gauge/gravity correspondence, Mateos {\it et al.}~studied the phase transition of quark-antiquark bound states (mesons). Their model was very similar to Frolov's toy model, and using the results of \cite{Frolov}, they demonstrated that in the case of an infinitely thin brane, the system generally undergoes a first order phase transition characterized by a change in the meson spectrum. The corresponding phase diagram in the vicinity of the critical solution exhibits a self similar structure, and this critical behavior and the first order transition are essentially universal to all Dp/Dq systems.

In \cite{MMT2} it was also pointed out, that higher order corrections to the brane effective action may cause, in principle, modifications to the above picture and it is likely 
that they spoil the system's scaling symmetry and self-similar behavior. Indeed, higher-derivative corrections to the D-brane action correspond to finite 't Hooft coupling corrections in the holographic dual, and provide a more realistic description of the system. These corrections may become important in the vicinity of the phase transition, since the curvature of the brane becomes large there. 

In the context of low-scale gravity theories, the possibility that a micro black hole may form in high energy collisions, like those at the LHC, re-creates a setup similar to the BBH system described above. In particular, the question whether a black hole may escape into the extra dimensional bulk has raised some attention, due to the potential phenomenological relevance \cite{flactan}. Clarifying the role of the thickness of the brane in that context is also an important  issue. 

The dynamics of branes keeps also attracting attention in the context of higher dimensional generalizations of the Bernstein conjecture \cite{B1,B2} and the study of the stability of brane-black hole systems \cite{B3}.

For all the above reasons, it is important to go beyond the approximation of zero thickness and consider higher order, curvature corrections coming from small thickness perturbations in the BBH system. Curvature corrections to the dynamics of domain walls without self-gravitation, in the case of non-zero thickness have been investigated earlier by Carter and Gregory \cite{Carter, Carter2}. They demonstrated that the next to leading order contribution is quadratic in the wall width (the brane thickness) and they obtained an exact, analytic expression for the corresponding effective action in terms of the intrinsic Ricci scalar $R$ and the extrinsic curvature scalar $K$. 

In a recent paper \cite{FG}, Frolov and Gorbonos studied the role of curvature corrections on topology changing transitions based on the effective brane action presented in \cite{Carter, Carter2}. In this work the authors focused their attention to the near-critical solutions and similarly to \cite{Frolov}, they restrict their investigations to the Rindler zone, i.e.~very close to the black hole horizon, where the radius of the intersection of the brane with the bulk horizon is much smaller than the radius of the bulk horizon. As an interesting result they found that ``the second order phase transition in such a system is modified and becomes first order''. This, however, seems to be in contradiction with the results of Mateos {\it et al.}~(see e.g.~\cite{MMT1}), where the authors demonstrated that the phase transition in the unperturbed system is generally a first order one. Additionally, they find that when the spatial dimension of the brane is larger than 2, supercritical solutions behave quite differently from subcritical ones, and for supercritical solutions there is no singularity resolution. According to their numerical analysis, they did not find evidence for the existence of such solutions. A possible explanation for this is that stiffness correction to the brane action break the symmetry between the super- and subcritical solutions

In the present paper we re-consider thickness perturbations to the BBH system. We proceed within a more general framework than that of \cite{FG}. We consider the same curvature corrected effective brane action obtained by Carter and Gregory \cite{Carter, Carter2}, but we do not restrict ourselves to work in the Rindler zone and in the near-critical solution region. In addition, we choose to follow a different path to obtain the dynamical equation for the perturbations. 

The plan of this paper is as follows. In Sec.~II we make a quick overview of the model in the infinitely thin case, and reintroduce the BBH setup to make the paper self-contained. In Sec.~III the curvature quantities are discussed, while in Sec.~IV we obtain the Euler-Lagrange equation for the curvature corrected brane action. As for the latter, we follow the perturbative treatment by Carter and Gregory \cite{Carter2}, in the sense that we treat the curvature corrections as small perturbations in the effective action. They are indeed very small as being quadratic in the perturbation parameter. To obtain the dynamical equation for the perturbations, we use the quadratic perturbation parameter to expand the $4th$-order Euler-Lagrange equation and keep the linear terms only. As a result, a second order, linear equation is found to describe the perturbation function $\vp$, with a very complicated source term. In Sec.~V we analyze the asymptotics of the perturbation equation, and find that there 
is no regular subcritical solution on the axis of the system above a certain dimension. This implies that our perturbation method is not appropriate in this region and a non-perturbative solution may be in order to find the general solution of the problem. In sec.~VI we write down the full Euler-Lagrange equation, but due to its very complicated and highly nonlinear form we do not discuss its solution in the present paper. In Sec.~VII we present the analytic solution for far distances and the numerical solution in the near horizon region of the perturbation equation for various dimensions. In Sec.~VIII we address the question of the phase transition in the case of a $D=3$ dimensional brane. For this purpose we use the approach of Flachi {\it et al.}~\cite{Flachi} based on the energy properties of a quasi-static brane evolution from the equatorial configuration. 

\section{The thin brane model}\label{s2}

Let us overview, in this section, the important properties of the BBH system, introduced in \cite{Frolov}, that we intend to study in the presence of a small brane thickness in the following sections. We consider static brane configurations in the background of a static, spherically symmetric bulk black hole. The metric of an $N$-dimensional, spherically symmetric black hole spacetime is
\begin{equation}
ds^2=g_{ab}dx^adx^b=-fdt^2+f^{-1}dr^2+r^2d\Omega_{N-2}^2\ ,
\end{equation}
where $f=f(r)$ and $d\Omega_{N-2}^2$ is the metric of an $N-2$ dimensional unit sphere. One can define coordinates $\theta_i (i=1,\dots, N-2)$ on this sphere with the relation
\begin{equation}
d\Omega_{i+1}^2=d\theta_{i+1}^2+\sin^2\theta_{i+1} d\Omega_i^2 \ .
\end{equation}
The explicit form of $f$ is not important, it is only assumed that $f$ is zero at the horizon $r_0$, and it grows monotonically to $1$ at the spatial infinity $r\rightarrow \infty$, 
where it has the asymptotic form \cite{Tangherlini},
\begin{equation}
f=1-\left(\frac{r_0}{r}\right)^{N-3}\ .
\end{equation}

In the zero thickness case, the test brane configurations, in an external gravitational field, can be obtained by solving the equation of motion coming from the Dirac-Nambu-Goto action \cite{Dirac, Nambu, Goto},
\begin{equation}\label{action0}
S=\int d^D\zeta\sqrt{-\mbox{det}\gamma_{\mu\nu}}\ , 
\end{equation}
where $\gamma_{\mu\nu}$ is the induced metric on the brane
\begin{equation}
\gamma_{\mu\nu} =g_{ab}\frac{\partial x^a}{\partial \zeta^{\mu}}
\frac{\partial x^b}{\partial \zeta^{\nu}}\ ,
\end{equation}
and $\zeta^{\mu}(\mu=0,\dots ,D-1)$ are coordinates on the brane world sheet. The brane tension does not enter into the brane equations, thus for simplicity it can be put equal to $1$. It is also assumed that the brane is static and spherically symmetric, and its surface is chosen to obey the equations
\begin{equation}
\theta_D=\dots =\theta_{N-2}=\pi/2\ .
\end{equation}
With the above symmetry properties the brane world sheet can be defined by the function $\theta_{D-1}=\theta(r)$ and we shall use coordinates $\zeta^{\mu}$ on the brane as
\begin{equation}
\zeta^{\mu}=\{t,r,\phi_1,\dots,\phi_{n}\}\quad \mbox{with} \quad n=D-2 \ . 
\end{equation}
With this parametrization the induced metric on the brane is
\begin{equation}
\gamma_{\mu\nu} d\zeta^{\mu}d\zeta^{\nu}=-fdt^2+\left[\frac{1}{f}+r^2{\dot\theta}^2\right]dr^2+r^2\sin^2\theta d\Omega_n^2,
\end{equation}
where, and throughout this paper, a dot denotes the derivative with respect to $r$, and the action (\ref{action0}) reduces to
\begin{eqnarray}
S&=&\Delta t \mathcal{A}_n\int\mathcal{L}_0\ dr\ ,\\
\mathcal{L}_0&=&r^n\sin^n\theta\sqrt{1+fr^2{\dot\theta}^2}\ ,\label{L0} 
\end{eqnarray}
where $\Delta t$ is the interval of time and $\mathcal{A}_n=2\pi^{n/2}/\Gamma(n/2)$ is the surface area of a unit $n$-dimensional sphere.

The brane configurations are the solutions of the Euler-Lagrange equation
\begin{equation}
\frac{d}{dr}\left(\frac{\partial \mathcal{L}_0}{\partial \dot\theta}\right)-\frac{\partial \mathcal{L}_0}{\partial \theta}=0\ ,
\end{equation}
which for the Lagrangian (\ref{L0}) reads
\begin{equation}\label{eq0}
\ddot\theta + B_3{\dot\theta}^3+B_2{\dot\theta}^2+B_1{\dot\theta}+B_0=0\ ,
\end{equation}
\begin{eqnarray}\label{coeff0} 
B_0&=&-\frac{n\cot\theta}{fr^2},\quad B_1=\frac{n+2}{r}+\frac{\dot f}{f},\nonumber\\
B_2&=&-n\cot\theta, \quad B_3=r\left[\frac{r\dot f}{2}+(n+1)f\right]\ .
\end{eqnarray}
For the supercritical case, a regular solution of (\ref{eq0}) has the expansion near the 
horizon 
\begin{equation}\label{h-expand}
\theta=\theta_0+\dot\theta_0(r-r_0)+\dots,\quad\mbox{with}\quad 
\dot\theta_0=\left.\frac{n\cot\theta}{\dot fr^2}\right|_{r_0},
\end{equation}
hence it is uniquely determined by the initial condition $\theta_0$. In the subcritical case the brane does not cross the horizon, and its surface reaches the minimal distance from the black hole at $r_1>r_0$ which, for symmetry reasons, occurs at $\theta=0$. A regular solution of (\ref{eq0}) near this point has the behavior 
\begin{equation}\label{o-expand}
\theta=\eta\sqrt{r-r_1}+\sigma(r-r_1)^{3/2}+\dots\ ,
\end{equation}
where
\begin{eqnarray}
\eta&=&\left.\sqrt{\frac{2(n+1)}{B_3(r)}}\right|_{r_1},\\ 
\sigma&=&\tfrac{1}{48(n+3)\eta r^2f}\Bigl[48n-\eta^2r\Bigl[6\eta^2(n+1)rf^2\nonumber\\
&+&24r\dot f +f\Bigl[4(n(r\eta^2+6)+12)\nonumber\\
&+&3r^2\eta^2(2(n+2)\dot f+r\ddot f)\Bigr]\Bigr]\Bigr]_{r_1}
\end{eqnarray}
and hence the solution is uniquely determined by the parameter $r_1$. In \cite{Frolov} the coefficient $\sigma$ is not considered, however it is necessary to fix its value to get a regular solution of (\ref{eq0}), and it also appears later on in the perturbation equation. 
\begin{figure}[!ht]\label{zeropic}
\noindent\hfil\includegraphics[scale=1]{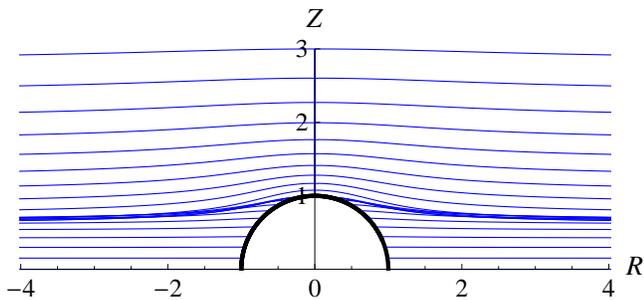} 
\caption{The picture shows a sequence of subcritical and supercritical thin brane solutions 
in the case when a $D=4$ dimensional brane embedded into a $N=5$ dimensional bulk. The different configurations belong to different initial conditions of $r_1$ and $\theta_0$, respectively. For simplicity, the bulk black hole's horizon radius is put to be 1, and $R$ and $Z$ are the standard cylindrical coordinates.}
\end{figure}

The far distance solution of (\ref{eq0}) can be obtained from the condition that the brane behaves asymptotically as a $(D-1)$-dimensional plane. The solution can be searched in the form
\begin{equation}\label{inftheta}
\theta=\frac{\pi}{2}+\nu(r)\ ,
\end{equation}
and one can get for $\nu(r)$ 
\begin{equation}\label{nu}
\nu(r)=\left\{ 
\begin{array}{cc}
\frac{p}{r}+\frac{p'}{r^n} & \mbox{if $n>1$,}\\
\frac{p+p'\ln r}{r} & \mbox{if $n=1$,}
\end{array}\right.
\end{equation}
where $p$ can be referred as the distance of the brane from the equatorial plane at infinity, and both $p$ and $p'$ are well defined continuous functions of the initial parameters $\theta_0$ or $r_1$.

\section{Curvature corrections}
Small thickness perturbations to the brane dynamics are derived from higher-order, 
curvature corrections to the effective action of the brane. In the present model we do not consider the self-gravitation of the brane, hence the curvature scalars are completely determined by the embedding black hole spacetime. In the approximation when all the relevant dynamical length scales $L$ of the system are very large compared to the parameter $\ell$ that characterizes the thickness of the brane, Carter and Gregory obtained the following exact, analytic expression for the effective action of the brane dynamics \cite{Carter2},  
\begin{equation}\label{action1}
\!\!S=\int d^D\zeta\sqrt{-\mbox{det}\gamma_{\mu\nu}}\left[-\frac{8\mu^2}
{3 \ell}(1+C_1R+C_2K^2)\right],
\end{equation}
where $R$ is the Ricci scalar, $K$ is the extrinsic curvature scalar and the coefficients $C_1$ and $C_2$ are expressed by the wall thickness parameter as
\begin{equation}
C_1=\frac{\pi^2-6}{24}\ell^2\ , \qquad C_2= -\frac{1}{3}\ell^2 .
\end{equation}
The parameter $\mu$ is related to the thickness as 
\begin{equation}
\ell=\frac{1}{\mu\sqrt{2\lambda}}
\end{equation}
which originates from a field theoretical domain-wall model, where $\mu$ is the mass parameter and $\lambda$ is the coupling constant of the scalar field.

Now let us consider the $N$ dimensional BBH setup described in Sec.~\ref{s2}, where the thickness corrected effective action (\ref{action1}) reduces to 
\begin{eqnarray}\label{S}
S&=&\Delta t \mathcal{A}_n\int\mathcal{L}\ dr\ ,\\
\mathcal{L}&=&-\frac{8\mu^2}{3\ell}\mathcal{L}_0[1+\ve\delta]\ ,\label{L} 
\end{eqnarray}
where we introduced the notations 
\begin{eqnarray}\label{delta}
\ve=\frac{\ell^2}{L^2},\qquad \delta&=&\left[\frac{C_1L^2}{\ell^2}R+\frac{C_2L^2}{\ell^2}K^2\right]\ .
\end{eqnarray}
In order to calculate the curvature scalars we follow the method described in \cite{Gregory, Carter2} using the Gauss-Codazzi formalism to split the quantities into their components orthogonal and parallel to the brane world sheet. 

Let $n^a$ be a unit geodesic vector field orthogonal to the brane and $z$ the length parameter along the integral curves of $n^a$. Then each constant $z$ hypersurface has an intrinsic metric $h_{ab}$ and extrinsic curvature $K_{ab}$ defined by
\begin{eqnarray}
n_a&=&\frac{\nabla_az}{\sqrt{g^{ab}\nabla_az\nabla_bz}},\\
h_{ab}&=&g_{ab}-n_an_b,\\
K_{ab}&=&h_a^{c}\nabla_cn_b,
\end{eqnarray}
where the length parameter in our chosen spherical coordinate system is
\begin{equation}
z=\int \sqrt{g_{ab}\tfrac{dx^a}{d \tau}\tfrac{dx^b}{d \tau}}d \tau
\equiv\int \sqrt{\tfrac{{r'}^2}{f}+r^2{\theta'}^2}d \tau,
\end{equation}
$\tau$ is the curve parameter and the prime denotes the derivative with respect to $\tau$.
We make the assumption that the perturbed system preserves the original symmetry properties 
of the unperturbed BBH setup, hence expressing $z$ as a function of $r$ we find that the only 
non vanishing components of $n^a$ are
\begin{eqnarray}
n^1=\frac{fr\dot\theta}{\sqrt{1+fr^2{\dot\theta}^2}},\quad n^2&=&-\frac{1}{r\sqrt{1+fr^2{\dot\theta}^2}}\ .
\end{eqnarray}
The extrinsic curvature scalar can be obtained as
\begin{eqnarray}
K\equiv K_{a}^a=h^{c}_a\nabla_cn^a\ ,
\end{eqnarray}
and the Ricci scalar of the intrinsic metric $\gamma_{ab}$ of the brane world sheet is given by the Gauss formula
\begin{equation}
R=K^2-K^a_bK^b_a\equiv K^2-Q\ ,\label{gauss}
\end{equation}
where we introduced the notation $Q$ for $K^a_bK^b_a$. Now we can rewrite $\delta$ of (\ref{delta}) as
\begin{equation}\label{abdelta}
\delta=aK^2+bQ
\end{equation}
with
\begin{equation}\label{ab}
a=\frac{\pi^2-14}{24}L^2,\quad b=\frac{6-\pi^2}{24}L^2\ ,
\end{equation}
and the curvature scalars $K$ and $Q$ are
\begin{eqnarray}\label{K}
&&\!\!\!\!\!\!\!\!\!\!\!\!\!\!\!\!\!\!\!\!
K=\frac{1}{F}\left[\frac{r\dot\theta\dot f}{2}+\frac{B}{2F^2}+(n+1)f\dot\theta-\frac{n \cot\theta}{r} \right],\\ 
&&\!\!\!\!\!\!\!\!\!\!\!\!\!\!\!\!\!\!\!\!
Q=\frac{1}{F^2}\left[\frac{r^2{\dot\theta}^2{\dot f}^2}{4}+\frac{B^2}{4F^4} +\frac{f\dot\theta B}{F^2}\right.\nonumber\\
&&\left.\quad+f^2{\dot\theta}^2 +n\left[f\dot\theta-\frac{\cot\theta}{r}\right]^2 \right] , \label{Q}
\end{eqnarray}
where
\begin{eqnarray}\label{AB}
F&=&\sqrt{1+fr^2{\dot\theta}^2}\ , \\
B&=&\left[r\dot f+2f\right] \dot\theta+4rf\ddot\theta\ . \label{B}
\end{eqnarray}

With the obtained expressions for the curvature scalars we will write up the Euler-Lagrange equation of the curvature corrected problem, and look for its regular solutions within a perturbative approach.

\section{The brane equation}
Since the thickness corrected effective action (\ref{action1}) also depends on the second derivative of $\theta$ (as one can expect it from the presence of the curvature terms), 
the dynamics of the perturbed brane is described by the 4$th$-order Euler-Lagrange
equation,
\begin{equation}\label{mel}
\frac{d^2}{dr^2}\left(\frac{\partial \mathcal{L}}{\partial \ddot\theta}\right)
-\frac{d}{dr}\left(\frac{\partial \mathcal{L}}{\partial \dot\theta}\right)+\frac{\partial \mathcal{L}}{\partial \theta}=0\ .
\end{equation}
Plugging the effective Lagrangian (\ref{L}) into (\ref{mel}) we get
\begin{eqnarray}\label{lagp}
\!\!\!\!\!\!\!\!\!\!\!\!\!\!\!\!\!\!\!\!
0&=&\frac{d}{dr}\left(\frac{\partial \mathcal{L}_0}{\partial \dot\theta}\right)-\frac{\partial \mathcal{L}_0}{\partial \theta}\nonumber\\
\!\!\!\!\!\!\!\!\!
&-&\ve\left[\frac{d^2}{dr^2}\left(\frac{\partial (\mathcal{L}_0\delta)}{\partial \ddot\theta}\right)-\frac{d}{dr}\left(\frac{\partial (\mathcal{L}_0\delta)}{\partial \dot\theta}\right)+\frac{\partial (\mathcal{L}_0\delta)}{\partial \theta}\right].
\end{eqnarray}

At this point, to solve (\ref{lagp}), we choose a different path 
than the one that was followed by Frolov and Gorbonos in \cite{FG}. First, we do not restrict 
our investigations to the Rindler zone and/or to the near critical solution of the thin problem; second, we follow more closely the treatment of Carter and Gregory in \cite{Carter2}, in the sense that, since the curvature corrections in (\ref{S}) are second order in the perturbation parameter $\ell/L$, we are looking for a perturbed solution whose perturbation is also quadratic in the same parameter. Hence we shall proceed by expressing the solution of (\ref{lagp}) as a power series in $\ve=\ell^2/L^2$ as
\begin{equation}\label{exp}
\tilde\theta(r)=\theta(r)+\ve \vp(r)+o(\ve^2),
\end{equation}
where $\theta$ is the solution of the zero thickness case, $\vp$ describes its thickness correction at order of $\ve$, and we neglect all the higher order terms. 

Since the curvature scalars get larger and larger as we get closer to the singularity, the corresponding dynamical length scales become smaller and smaller. In the case of a fixed brane thickness $\ell$, this implies that the perturbation parameter $\ve$ becomes larger too. To ensure the validity of the present perturbation method, we have to guarantee the condition that $\ve\vp(r)\ll\theta(r)$, with the magnitude of 
\[
 \left|\frac{\ve\vp}{\theta}\right|\lesssim 0.01\ ,
\]
with respect to the second order perturbative treatment. As a consequence, for every fixed brane thickness value $\ell$, there is a minimum length scale parameter $L$ that defines a condition on how close we can get to the singularity before the validity of the present perturbation formalism breaks down.  

By reducing the brane thickness $\ell$, of course, one can approach the singularity arbitrarily close. In this work however, we will chose a different approach. We fix the value of the perturbation parameter $\ve=10^{-2}$. This means that we consider the thickest brane configurations that are allowed by the present perturbation method. This choice imposes the condition
\[
\left|\frac{\vp}{\theta}\right|\lesssim 1\ , 
\]
and since $0\leq\theta\leq\tfrac{\pi}{2}$, the corresponding condition on the amplitude of the perturbation function is
\[
|\vp|\leq\frac{\pi}{2}\ .
\]
The above requirement tells us how close we can go to the singularity (i.e.~how 
small the dynamical length scale $L$ can be) before the present perturbation method breaks down for the case of the thickest possible brane configurations.

Let us now substitute (\ref{exp}) into (\ref{lagp}), and after performing the series expansion up to $\ve$, we obtain the following linear, inhomogeneous equation for the perturbation 
\begin{equation}\label{fieq}
\ddot\varphi+q_1\dot\varphi+q_0\varphi+q=0,
\end{equation}
where 
\begin{eqnarray}
q_1&=&B_1+2\dot\theta B_2+3\dot\theta^2B_3,\\
q_0&=&\frac{n}{\sin^2\theta}\left[\frac{1}{r^2f}+{\dot\theta}^2\right],\\
q&=&\tfrac{1}{8r^4fF^6}
\left[A_0\theta^{(4)}+A_1\theta^{(3)}+A_2\ddot\theta^3+A_3\ddot\theta^2+A_4f^{(3)} 
\right.\nonumber\\
&+&\left.A_5\ddot f+A_6f^4+A_7f^3+A_8f^2+A_9f+A_{10}\right].
\end{eqnarray}
The coefficient functions $B_i$ are the same of (\ref{coeff0}), while the coefficient functions $A_i$ are given in the Appendix.

Hence, the perturbed problem reduces to solving a second-order, linear equation for $\vp$ with 
a very complicated source term. In the next section we examine the asymptotic behavior and the regularity conditions of (\ref{fieq}) near the horizon, and in Sec.~VII we discuss its solution.

\section{Regularity conditions}\label{regcon}

For a brane crossing the horizon (black hole embedding branch), the unperturbed solution $\theta$ has the expansion (\ref{h-expand}) near $r_0$. In this limit the source $q$ has an $\alpha(r)/f$ singular behavior as $r\rightarrow r_0$ with a regular $\alpha_0\equiv \alpha(r_0)$. The explicit expression for $\alpha$ is given in the Appendix. Hence, in the black hole embedding (supercritical) case the perturbation equation (\ref{fieq}) has a regular singular point on the horizon. Requiring the regularity for $\vp$ at the horizon implies the boundary condition 
\begin{equation}
\dot\vp_0=-\frac{1}{m}\left[\frac{n\varphi_0}{\sin^2\theta_0}+\alpha_0\right]\ ,
\end{equation}
where $m=N-3$.  Thus the perturbations for the supercritical solutions are uniquely determined 
by the initial value $\vp_0$. For the later purpose of making a direct comparison with the thin solution, we shall require the same boundary conditions for the perturbed and unperturbed brane solutions, therefore we choose $\vp_0$ to be zero and hence we get
\begin{equation}
\dot\vp_0=-\frac{\alpha_0}{m}\ .
\end{equation}

If the brane does not cross the horizon (Minkowski embedding branch) the brane surface reaches its minimal distance to the black hole at $r=r_1>r_0$. Near this point $\theta$ has the asymptotic form (\ref{o-expand}), and one can find that $q$ has a singular behavior  
\begin{eqnarray}
q&\sim&\frac{c_1}{\sqrt{r-r_1}}+\frac{c_3}{(r-r_1)^{3/2}}
+\frac{c_5}{(r-r_1)^{5/2}},
\end{eqnarray}
as $r\rightarrow r_1$, where
\begin{equation}\label{c5}
c_5=\tfrac{n(n-1)\eta}{8(n+1)r_1}\left[2(a+2b)(n+1)f+(a+3b)r\dot f\right]_{r_1}\;,
\end{equation}
and the explicit forms of $c_1$ and $c_3$ are given in the Appendix. Thus the perturbation equation (\ref{fieq}), in the subcritical case, near the vertical axis ($\theta=0$) has the asymptotic form
\begin{eqnarray}\label{fir1}
&&\!\!\!\!\!\!\!\!\!\!\!
\ddot\vp+\frac{n+3}{2(r-r_1)}\dot\vp+\left[\frac{n}{4(r-r_1)^2}
+\frac{\xi}{r-r_1}\right]\vp\nonumber\\
&&\!\!\!\!\!\!\!\!\!\!\!
+\frac{c_1}{\sqrt{r-r_1}}+\frac{c_3}{(r-r_1)^{3/2}}+\frac{c_5}{(r-r_1)^{5/2}}=0,
\end{eqnarray}
where  
\begin{eqnarray}
\xi=\frac{n}{r_1^2\eta^2f(r_1)}+\frac{n\eta^2}{12}+\frac{n\sigma}{\eta}\ .
\end{eqnarray} 

Equation (\ref{fir1}) shows that no regular solution of (\ref{fieq}) exists at finite $r_1$, unless $c_5$ disappears. From (\ref{c5}) we can see that there are two possible ways to satisfy this condition. One way is to choose the coefficients $a$ and $b$ in the appropriate way to make $c_5$ disappear. Even though in (\ref{ab}) we have already fixed the values of these parameters (we identified them with the constants that are obtained in the relativistic domain wall case in \cite{Carter2}), however, so far we have not used their explicit values. Thus we could still consider the general case with arbitrary $a$ and $b$. The problem with this way of regularization is that the constants $a$ and $b$ would be strongly dependent on the initial condition parameter $r_1$, and hence provide a highly unstable solution. 
 
The other way of making $c_5$ disappear is to choose the brane dimension parameter $n=0$ or $n=1$. Indeed, instead of the previous option, we will follow this direction in the regularization process of (\ref{fir1}) and keep the parameters $a$ and $b$ fixed as given in (\ref{ab}). This is also in agreement with what was found numerically in \cite{FG}. (It can also be expected that, for $n>1$, when additional symmetries in the angular directions are imposed, the $n=1$ case should be sufficient to discuss the brane configurations. Obviously, this is not the most general solution, and a different approach should be used to discuss the general case). The reason of the irregular behavior of the perturbations for $n>1$ is due to the asymmetry between the sub- and supercritical configurations induced by the curvature corrections. It also seems obvious that going beyond this approximation, within a perturbative approach, will not resolve the problem. In \cite{FG} the authors suggest that quantum corrections may cure the above pathological behavior. Here, we take the simpler viewpoint by recording the fact that since 
the thin solution is not smooth on the axis, the perturbation method we used breaks down near this region for configurations outside the horizon. This, however, does not mean that physically reasonable solutions do not exist, as the explicitly constructed field theoretical domain wall solutions clearly show \cite{Fl1,Fl2}. Hence, in the geometrical Dirac-Nambu-Goto approach, the thick solutions appear to deviate significantly from the thin ones near the axis of the system, and thus, our perturbative approach around the thin solution can not provide  regular thick solutions for $n>1$. In order to study the subcritical solutions and the thickness corrections to the phase transition in the general case, one needs to solve equation (\ref{lagp}) in a non-perturbative way. In the following section we write down the explicit Euler-Lagrange equation for the non-perturbative problem, but due to its very complicated form we do not discuss its solution in the present paper. We rather stay with our perturbative approach and complete its full analysis in the next coming sections.

In the following we provide the results for those cases where our perturbative approach leads to completely regular solutions: 1-dimensional string or a 2-dimensional sheet, where $n\equiv D-2$, and $D$ is the total number of dimensions of the brane spacetime. In both cases, from the requirement of regularity, we get the asymptotic behavior of $\vp(r)$ near the point $r_1$ as
\begin{equation}
\vp=\kappa \sqrt{r-r_1}+\rho(r-r_1)^{3/2}+\dots\ ,
\end{equation}
where
\begin{eqnarray}
\kappa=-\frac{2c_3}{n+1}\ ,\qquad \rho=-\frac{\kappa\xi +c_1}{n+3} \ .
\end{eqnarray}
Thus the perturbations for the subcritical solutions are uniquely determined by the value of the parameter $r_1$.

\section{The general Euler-Lagrange equation in the subcritical case}

In the previous section we discussed that the perturbation method (\ref{exp}) breaks down
around $\theta=0$ (the vertical axis) for cases $n>1$ in the Minkowski embeddings, and hence 
one has to consider the full $4th$-order Euler-Lagrange equation (\ref{lagp}) in searching for 
a regular solution of the curvature corrected problem. The equation of motion in this general case is a very complicated and highly nonlinear one, 
\begin{eqnarray}\label{fulleq}
\theta^{(4)}&+&T_1(\ddot\theta,\dot\theta,\theta,\dot f,f,r)\theta^{(3)}\nonumber\\
&+&T_2(\ddot\theta,\dot\theta,\theta,f^{(3)},\ddot f,\dot f,f,r)=0\ ,
\end{eqnarray}
where
\begin{eqnarray}\label{T1}
T_1&=&\frac{1}{rfF^2}\left[\right.
4r\dot f+f(2(2+n)+r\dot\theta(2n\cot\theta
\nonumber\\
&+& r(\dot\theta(2f(n-3+nr\cot\theta\dot\theta)-r\dot f)
\nonumber\\
&-& 10rf\ddot\theta)))\left. \right]\ ,
\end{eqnarray}
and $T_2$ is given in the Appendix. 

A detailed analysis of this equation, and a systematic, non-perturbative study to find its possible regular solutions would take us far from our essentially perturbative approach, and hence it lies beyond the scope of the present paper. Nevertheless, it does not seem completely impossible to perform this analysis, and to make a first step in this direction, we write up the asymptotic behavior of (\ref{fulleq}) near $\theta=0$. This equation (obtained by taking the series expansion of (\ref{fulleq}) around $\theta=0$ up to linear order) reads
\begin{equation}\label{theta0eq}
\frac{S_3}{\theta^3}+\frac{S_2}{\theta^2}+\frac{S_1}{\theta}+S_0+S\theta =0\ ,
\end{equation}
where the functions $S$, $S_0$, $S_1$, $S_2$, and $S_3$ are given in the Appendix. With taking a look on the explicit forms of the $S_i$ functions, one can see that equation (\ref{theta0eq}) is essentially just as complicated as (\ref{fulleq}), and hence we will not discuss its properties further here. It is important to note however, that the study of this problem and finding its regular solutions are crucial for the understanding of the complete picture behind the phase transition in the curvature corrected system. 

\section{The perturbed solution}

\subsection{Far distance solution}

As $r\rightarrow\infty$ we can use the asymptotic form of $\theta$ given in (\ref{inftheta}) and (\ref{nu}). In the case of supercritical embedding, if $n>1$ it is enough to consider the leading order $p/r$ term to obtain the asymptotic behavior of the perturbation equation (\ref{fieq}) and we get
\begin{equation}\label{inffin}
\ddot\vp+\frac{n+2}{r}\dot\vp+\frac{n}{r^2}\vp+\frac{E}{r^5}=0\ ,
\end{equation}
where
\begin{equation}
E=-4p\left[(a+2b)(n-2)^2-a(n-4)\right] .
\end{equation}
Equation (\ref{inffin}) can be integrated in a closed form, and its solution reads
\begin{equation}\label{inffisoln}
\vp=\frac{P}{r}+\frac{P'}{r^n}+\frac{E}{2(n-3)r^3}\ . 
\end{equation}


From (\ref{inffisoln}) we can see that in the case of $n=3$ the equation (\ref{inffin}) develops a resonance source term and hence (\ref{inffisoln}) is not 
a solution. In this special case the solution reads
\begin{equation}\label{inffisol3}
\vp=\frac{P}{r}+\frac{P'}{r^3}+\frac{E\left[1+2\ln r\right]}{4r^3}\ . 
\end{equation}

In the case of $n=1$ the perturbation equation (\ref{fieq}), at far distances takes the asymptotic form 
\begin{equation}\label{inffi1}
\ddot\vp+\frac{3}{r}\dot\vp+\frac{1}{r^2}\vp+\frac{E_1+E_2\ln r}{r^5}=0\ ,
\end{equation}
where
\begin{eqnarray}
E_1&=&4\left[a(11p'-4p)+b(7p'-2p)\right] ,\\
E_2&=&-8p'(2a+b) .
\end{eqnarray}
The solution of (\ref{inffi1}) is
\begin{equation}\label{inffisol1}
\vp=\frac{P+P'\ln r}{r}-\frac{E_1+E_2(1+\ln r)}{4r^3}\ ,
\end{equation}
and the integration constants $P$ and $P'$ in (\ref{inffisoln}), (\ref{inffisol3}) and (\ref{inffisol1}) are well defined continuous functions of $\vp_0$ or $r_1$.

\subsection{Perturbations near the horizon}
After the analysis of the regularity conditions and the asymptotic behavior of the perturbation equation near the event horizon, it is now fairly straightforward to solve (\ref{fieq}) numerically. There is, however, one last thing that we have to discuss before presenting the numerical results. In Sec.~IV we have already mentioned that the dynamical length scale of the problem has to be identified, in order to know how close one can get to the singularity before the present perturbative approach breaks down. Since 
\begin{equation}
\frac{1}{L} \sim \mbox{max}\{K,\sqrt{|R|}\}, 
\end{equation}
we can calculate $L$ for every initial parameter $\theta_0$ and $r_1$ from the supercritical and subcritical thin solutions respectively. We can then plug $L$ into the curvature coefficients $a$ and $b$ of (\ref{abdelta})-(\ref{ab}), and solve the perturbation equation. The requirement of how small $\theta_0$ can be, i.e.~how close we can get to the singularity before the quadratic perturbation approximation breaks down, is then obtained from the condition that $|\vp|\leq\frac{\pi}{2}$. This condition uniquely determines the length scale of the perturbation problem in every chosen bulk ($N$) and brane ($D$) dimensions.

In TABLE I we have listed the approximate minimal values of the initial parameters $\theta_0$ of those thick brane configurations that can still be addressed by the present perturbation method. These values are used in the numerical calculations to obtain the dynamical length scale of the perturbed system in various bulk and brane dimensions. It is interesting to notice that with increasing co-dimension ($N-D$) the minimum value of $\theta_0$ is decreasing. Another tendency is that keeping the co-dimension fixed while increasing the bulk dimension, the minimum $\theta_0$ is also increasing. 
\begin{center}
\begin{table}[ht]
\begin{tabular}{|c||c|c|c|c|c|}
        \hline
$D\diagdown N$ & $5$ & $6$ & $7$ & $8$ & $9$ \\
        \hline\hline
$3$  & $6.82^{\circ}$ & $4.52^{\circ}$ & $3.4^{\circ}$ & $2.7^{\circ}$ & $2.25^{\circ}$ \\
        \hline
$4$  & $10.29^{\circ}$ & $7.14^{\circ}$ & $5.43^{\circ}$ & $4.34^{\circ}$ & $3.64^{\circ}$ \\
        \hline
$5$  & - & $9.85^{\circ}$ & $7.56^{\circ}$ & $6.18^{\circ}$ & $5.14^{\circ}$\\
        \hline
$6$  & - & - & $10.12^{\circ}$ & $8.26^{\circ}$ & $6.96^{\circ}$ \\
        \hline
$7$  & - & - & - & $10.48^{\circ}$ & $8.82^{\circ}$\\
        \hline
\end{tabular}
\caption{The approximate values of the initial parameter $\theta_0$ for $N=5,\dots,9$ and $D=3,\dots,7$, where the present perturbation method reaches its limitation, i.e.~where $|\vp|\simeq \frac{\pi}{2}$.}
\end{table}
\end{center}
In the rest of this section we present the numerical solution of the perturbation equation
(\ref{fieq}) in various dimensions. First we consider the case when $n>1$, i.e.~when there 
is no regular subcritical perturbations. We analyze this type of solution in the case 
of $n=2$ with varying bulk dimensions. Later on we consider the $n=1$ case, both for the subcritical and supercritical solutions.

\subsubsection{The $n=2$, supercritical case.}
\begin{figure}[!ht]\label{fig:n=2_bhperts}
\noindent\hfil\includegraphics[scale=.5]{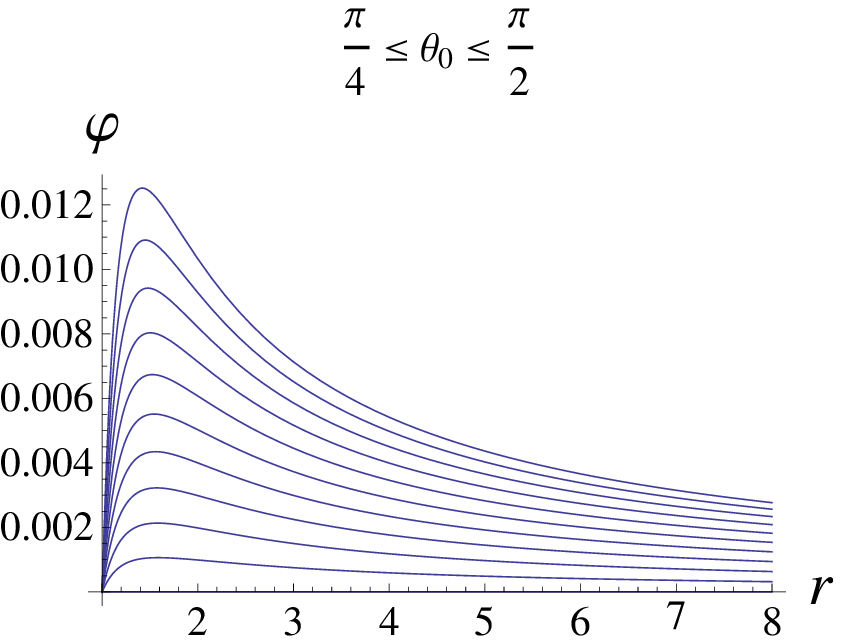} 
\noindent\hfil\includegraphics[scale=.5]{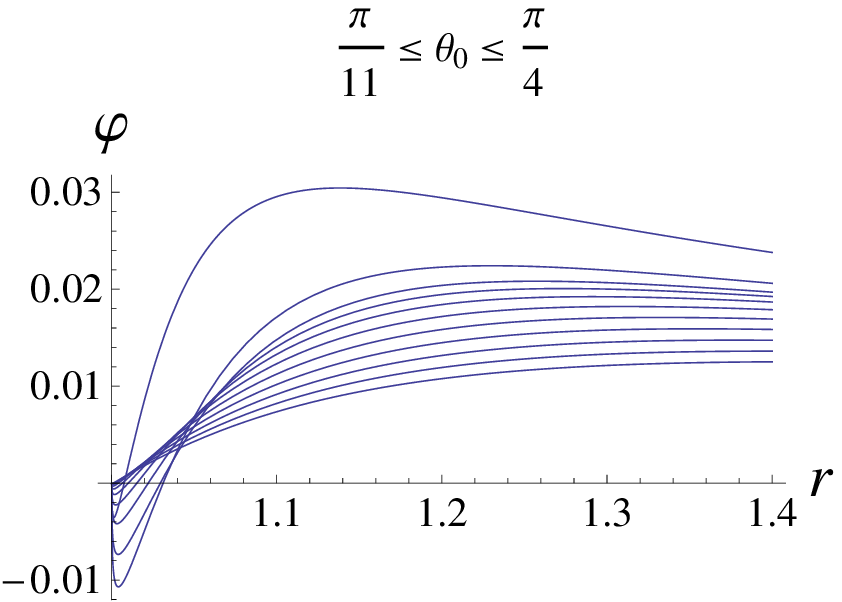} 
\vskip 5pt
\noindent\hfil\includegraphics[scale=.5]{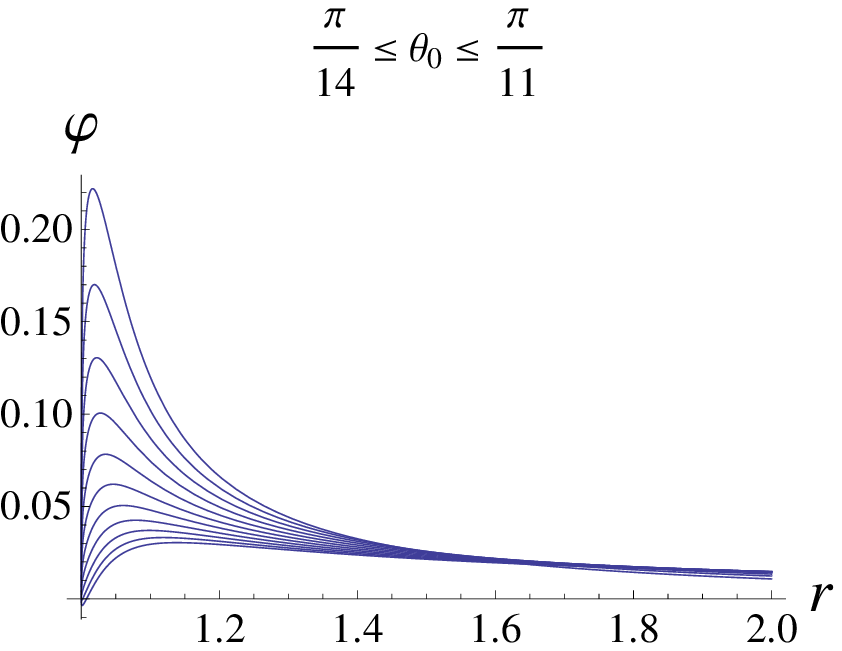} 
\noindent\hfil\includegraphics[scale=.5]{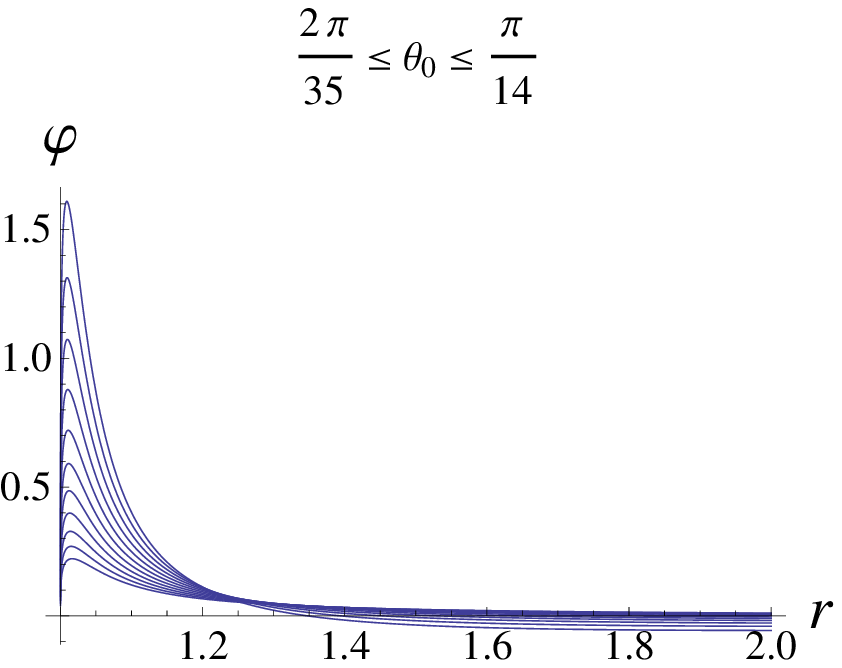}
\caption{The picture shows the numerical solution of the perturbation equation (\ref{fieq}) 
with varying the parameter $\theta_0$ of the brane initial inclination in the case of $N=5$, $n=2$, black hole embedding.}
\end{figure}
In FIG.~2 we have plotted the perturbations of a $D=4$ dimensional brane embedded into 
a $N=5$ dimensional bulk spacetime. By varying the initial parameter $\theta_0$ that 
describes the brane position on the horizon ($\theta$ measures the inclination from 
the vertical axis) until we get to its minimum, one can find four regions with qualitatively different behavior. In the first region (top left), the perturbations are positive from the horizon to infinity, and their maximum amplitude grows moderately 
as $\theta_0$ decreases. In the second region (top right), the perturbations are negative in the vicinity of the horizon, but soon become positive and after some increasing phase they start decaying. In the $3rd$ region (bottom left), the picture is similar to the 1st one with the difference that the amplitude starts growing more rapidly as $\theta_0$ decreases. Finally, in the $4th$ region (bottom right), the perturbations are positive in the vicinity of the horizon, and there is a sign change near the horizon before they start decaying. 
\begin{figure}[!ht]\label{fig:n=2_N=5678_bhperts}
\noindent\hfil
\includegraphics[scale=.5]{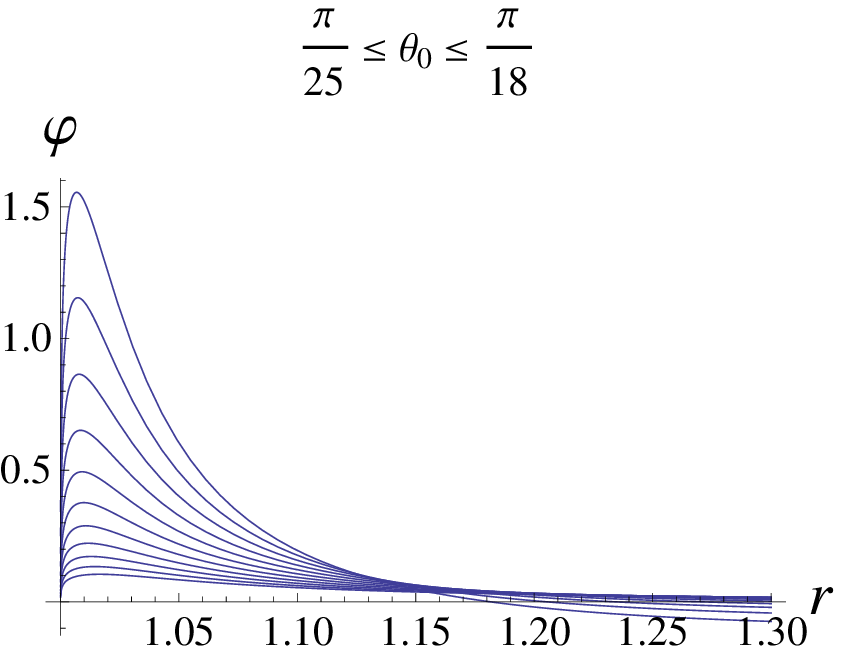} 
\includegraphics[scale=.5]{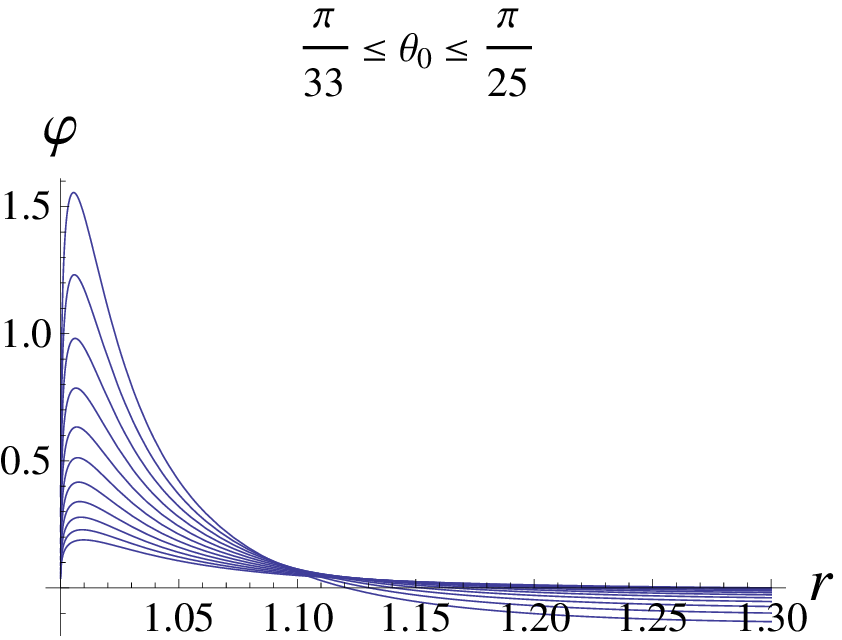}
\vskip 5pt\noindent\hfil
\includegraphics[scale=.5]{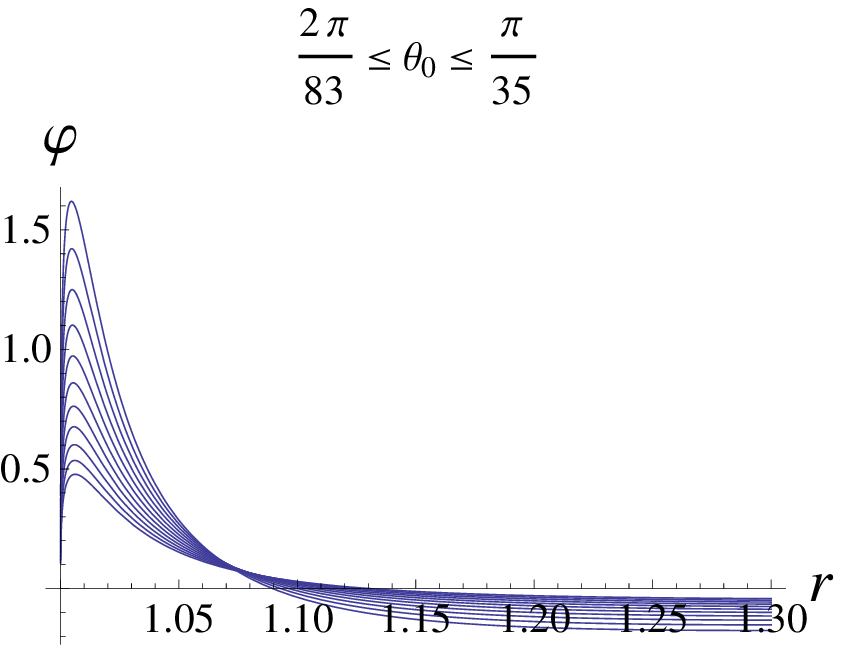} 
\includegraphics[scale=.5]{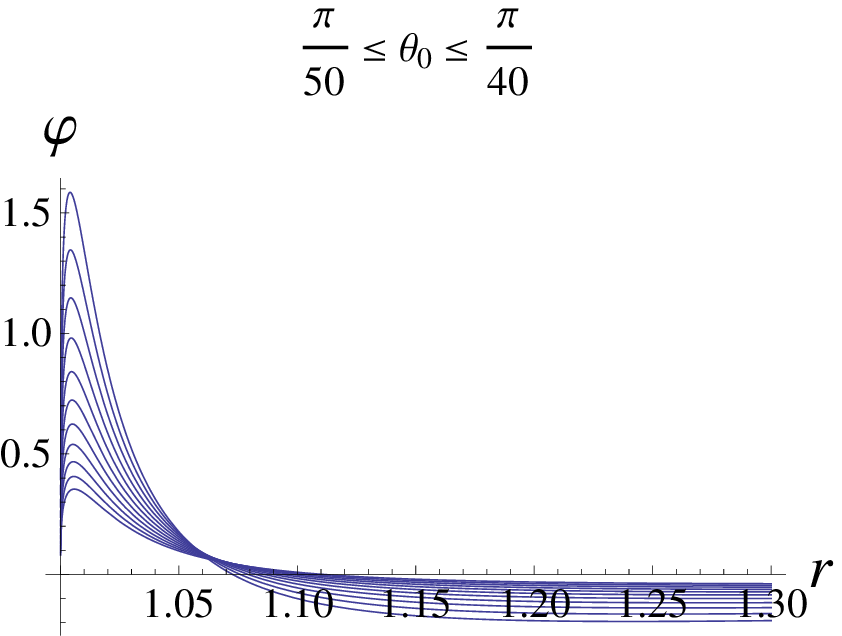}
\caption{The picture shows the numerical solutions of the perturbation equation (\ref{fieq}) 
in the case of a $D=4$ dimensional ($n=2$) brane embedded into a $N=6,7,8,9$ (top left, top right, bottom left, bottom right, respectively) dimensional bulk spacetime. The graphs belong 
to the initial conditions that are close to the minimal $\theta_0$, i.e.~when the perturbation method reaches its limit $|\vp|\simeq\frac{\pi}{2}$.}
\end{figure}

In FIG.~3 we have plotted the perturbations of the near minimal $\theta_0$ zone ($|\vp|\leq\frac{\pi}{2}$) in different bulk dimensions. By increasing the bulk dimension and keeping the brane dimension fixed, we found that the perturbations amplitude are decreasing and hence the minimal $\theta_0$ is also decreasing. This means that we can get closer to the singularity $\theta=0$ before the perturbation approximation breaks down, i.e.~$|\vp|\simeq\frac{\pi}{2}$. It is also interesting to notice that the sign change, in the near singularity region solutions, gets closer and closer to the event horizon with increasing bulk dimensions. 
\begin{figure}[!ht]\label{fig:N5n2bhpert}
\noindent\hfil\includegraphics[scale=1]{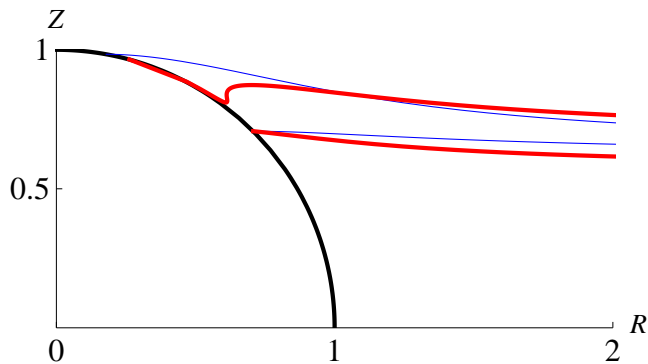} 
\caption{The picture shows the thick (red) brane configurations together with their thin (blue) counterparts in a cylindrical coordinate system, in the case of an $N=5$, $n=2$ black hole embedding. The initial conditions are $\theta_0=\tfrac{\pi}{4}$ (bottom curves) and $\tfrac{\pi}{17.5}$ (top curves), and the thickness parameter $\ell$ is chosen to be large for the purpose of making the effects visible. The black curve represents the black hole's event horizon.}
\end{figure}

In FIG.~4 we have plotted the corresponding thick (perturbed) and thin (unperturbed) brane configurations with the initial parameters $\theta_0=\tfrac{\pi}{4}$ (bottom curves) and $\tfrac{\pi}{17.5}$ (top curves) respectively, in a cylindrical coordinate system with $N=5$ bulk dimensions. On the picture, the black curve represents the black hole's event horizon, the blue thin curves the unperturbed brane solutions, while the red, thick curves the perturbed ones. 

\subsubsection{The $n=1$ case.}

In the case of $n=1$, i.e.~a $D=3$ dimensional brane, the perturbation equation (\ref{fieq}) has regular solutions both in the supercritical and the subcritical cases. For the black hole embedding branch the perturbations are very similar to the $n=2$ case. The only remarkable difference, compared to the results that are plotted in FIG.~2, is that in the $n=1$ case the $2nd$ region of FIG.~2 is missing, i.e.~the perturbations are always positive in the very vicinity of the horizon.

The numerical solutions of the perturbation equation in the case of the Minkowski embedding branch can be seen in FIG.~5. We have chosen two regions, characterized by their minimal distances from the black hole's event horizon. The first picture (top left) in FIG.~5 shows the perturbations in cases when the brane is very close to the horizon, $1.001\leq r_1\leq 1.01$. The second picture (top right), however, belongs to the solutions that are a bit further away from the horizon, in the region of $1.1\leq r_1\leq 3$. The top pictures belong to $N=5$, whilst the bottom pictures to $N=6$ embeddings. 
\begin{figure}[!ht]\label{fig:N5mpert}
\noindent\hfil\includegraphics[scale=.5]{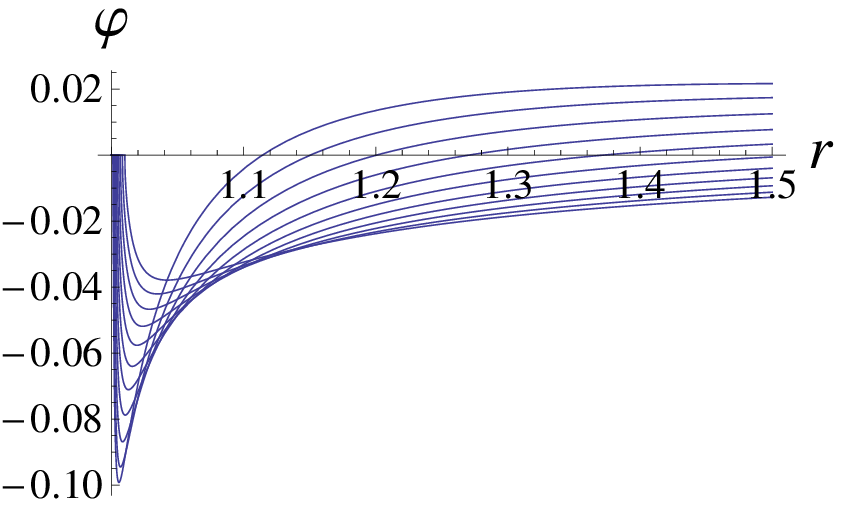} 
\noindent\hfil\includegraphics[scale=.5]{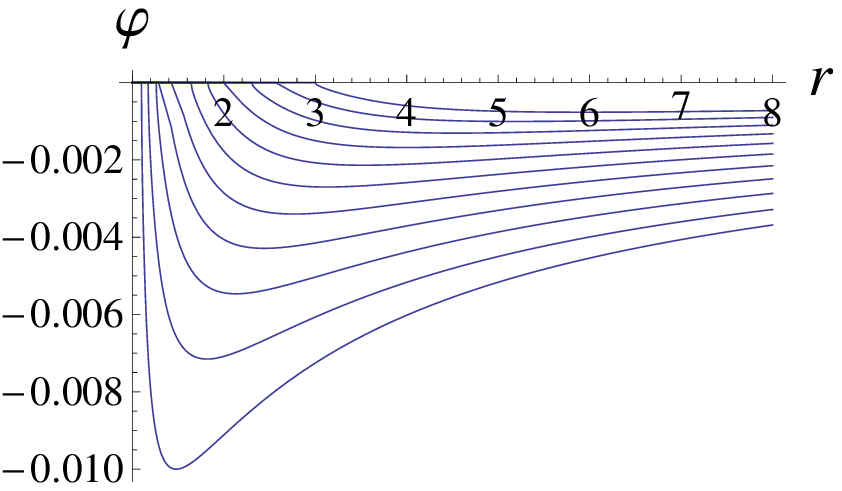} 
\noindent\hfil\includegraphics[scale=.5]{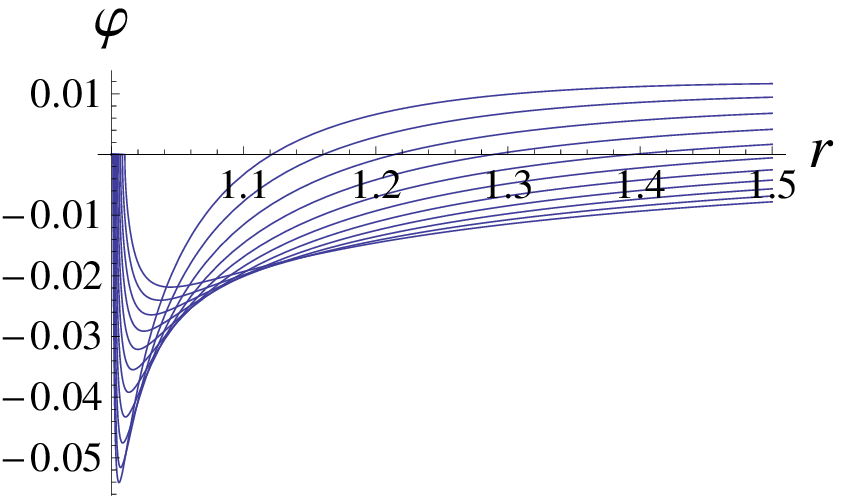} 
\noindent\hfil\includegraphics[scale=.5]{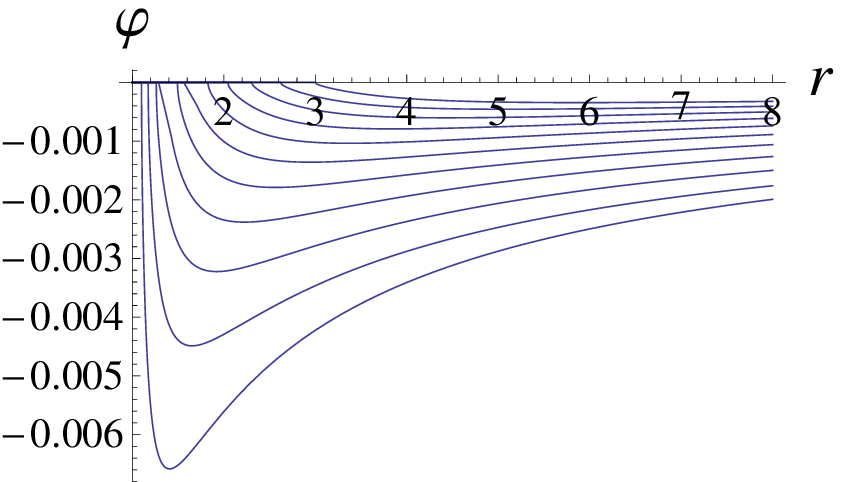} 
\caption{The figure shows the numerical solutions of the perturbation equation (\ref{fieq}) in 
the cases when a $D=3$ dimensional brane is embedded into an $N=5$ (top pictures) and an $N=6$ (bottom pictures) dimensional bulk. The pictures belong to the Minkowski embedding branch with initial parameter regions $1.001\leq r_1\leq 1.01$ (left hand side), and  $1.1\leq r_1\leq 3$, (right hand side).}
\end{figure}

From FIG.~5 we can see that the perturbations in the near horizon zone start with a negative maximum and soon after there is a sign change before they begin to decay. The far zone solutions start also with a negative maximum but they begin to decay without changing their sign. From the top to the bottom, we have plotted the same solutions changing the bulk dimension from 5 to 6. In accordance with the corresponding black hole embedding solutions, we can see that the perturbations' amplitudes are also decreasing with increasing bulk dimensions.

\begin{figure}[h]
\noindent\hfil\includegraphics[scale=1]{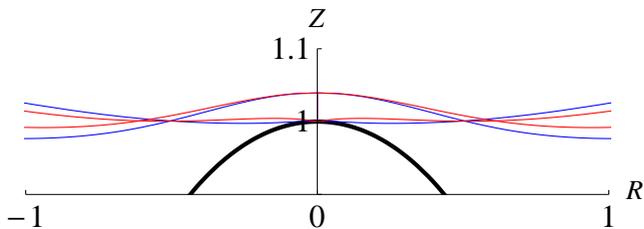} 
\caption{The picture shows the thick (red) brane configurations together with their thin (blue) counterparts in a cylindrical coordinate system, in the case of an $N=5$, $n=1$ Minkowski embedding. The initial parameters are $r_1=1.001$ (bottom curves) and $r_1=1.04$ (top curves), and the thickness parameter $\ell$ is chosen to be large for the purpose 
of making the effects visual. The black thick curve represents the black hole's event horizon.}
\end{figure}

In FIG.~6 we have plotted two corresponding thick (perturbed) and thin (unperturbed) brane configurations with initial parameters $r_1=1.001$ (bottom curves on the $Z$ axis) and $r_1=1.04$ 
(top curves on the $Z$ axis) respectively, in a cylindrical coordinate system with $N=5$ bulk dimensions. On the picture, the black curve represents the black hole's event horizon, the blue curves the unperturbed- (thin), while the red curves the perturbed (thick) brane solutions. We can see on the picture that the near horizon perturbed configuration intersects its thin counterpart as the perturbation changes its sign before it starts decaying. There is no intersection, however between the thick and thin configurations at the top curves, since the perturbations in the far zone do not change their sign before they start decaying. 

\section{Energy properties and phase transition}
In obtaining the solutions of the curvature corrected brane action (\ref{S}) with a second order perturbative approach in the thickness, we concluded that our perturbative method breaks down in the vicinity of the axis of the system, for the subcritical solutions with $n>1$, and hence one needs to find a new, exact solution for a general discussion. Nevertheless, we also found that in the special cases of $n=0$ and $n=1$, the perturbation equation (\ref{fieq}) can be regularly solved in the subcritical case too, whereas we also showed that the supercritical solutions are regular and can be solved without any difficulty for any $n$. In the following we will discuss the phase transition between the Minkowski embedding and black hole embedding topology of the thickness corrected brane solutions with $n=0$, $1$. Since we are interested in brane solutions we present the results of the $n=1$ case. 
\begin{figure}[!ht]\label{fig:energy0}
\noindent\hfil\includegraphics[scale=.5]{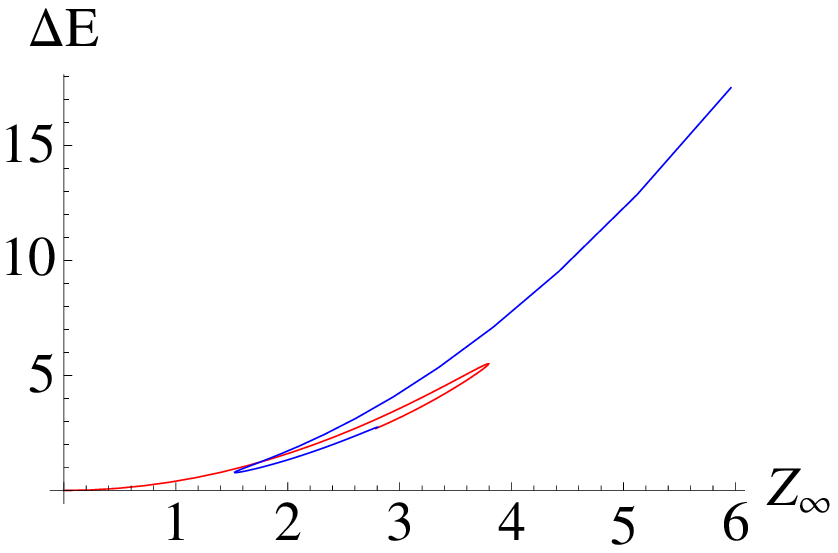} 
\noindent\hfil\includegraphics[scale=.5]{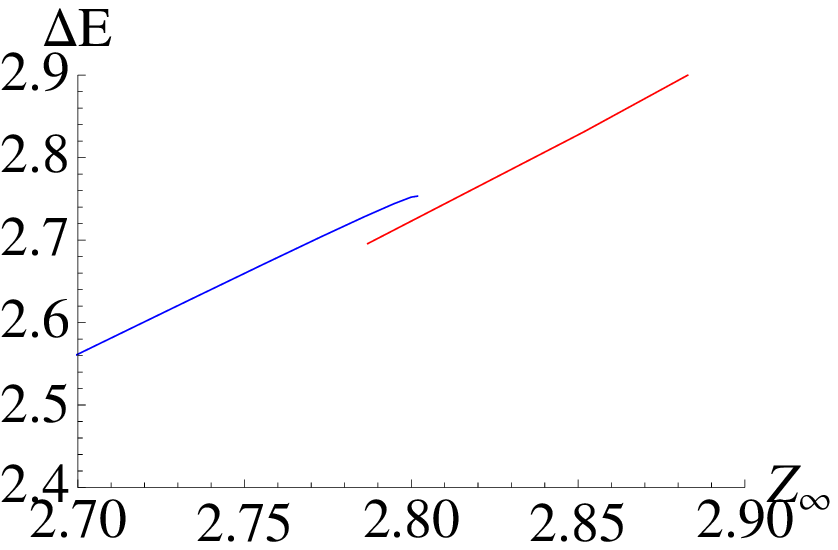} 
\noindent\hfil\includegraphics[scale=.5]{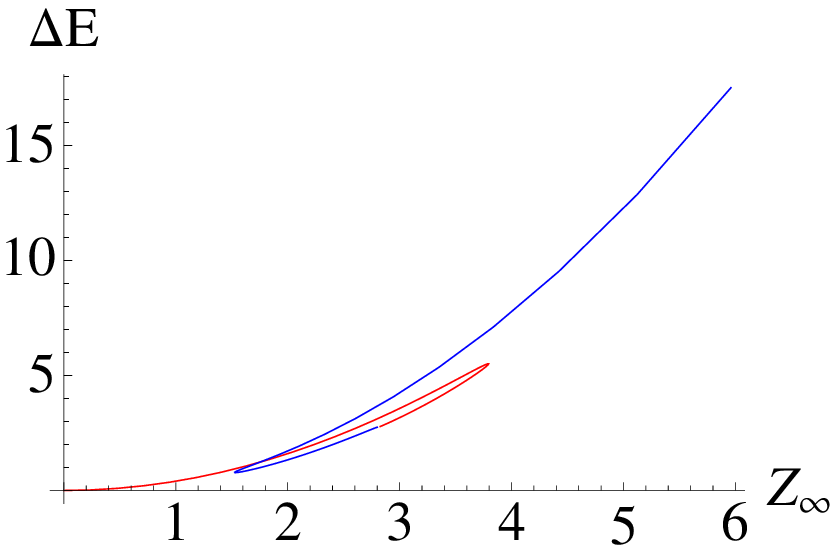} 
\noindent\hfil\includegraphics[scale=.5]{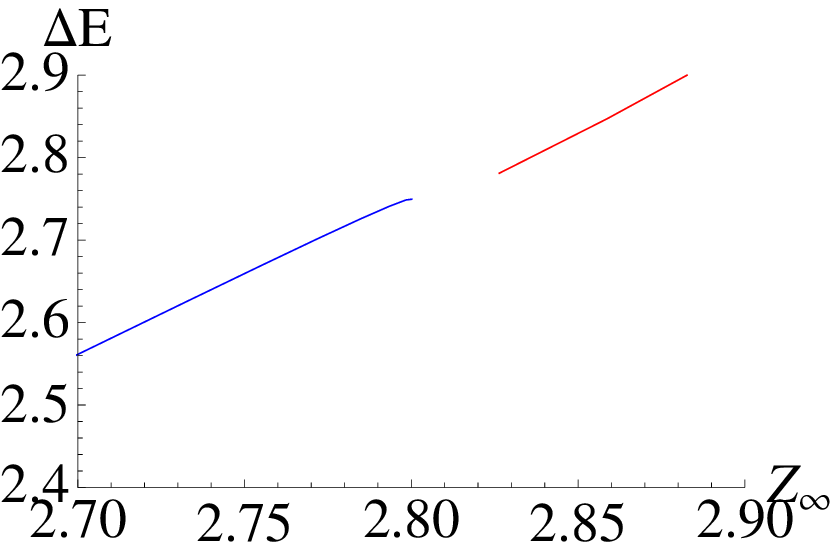} 
\caption{The figure shows the energy difference of a brane configuration (that quasi-statically evolves from the equatorial configuration) with respect to the equatorial configuration, as a function of the parameter $Z_{\infty}$ in the case of $N=5$ bulk dimensions. The top pictures belong to the thin, while the bottom pictures to the thick system. The red (blue) curves represent the black hole (Minkowskian) embedding branch. The pictures on the right hand side are the zooms into the near region of $\theta_{min}$, where the effects of the perturbations
are the largest, and the difference become apparent between the thin (top) and thick (bottom) cases.}
\end{figure}
\begin{figure}[!ht]\label{fig:energy_pert}
\noindent\hfil\includegraphics[scale=.5]{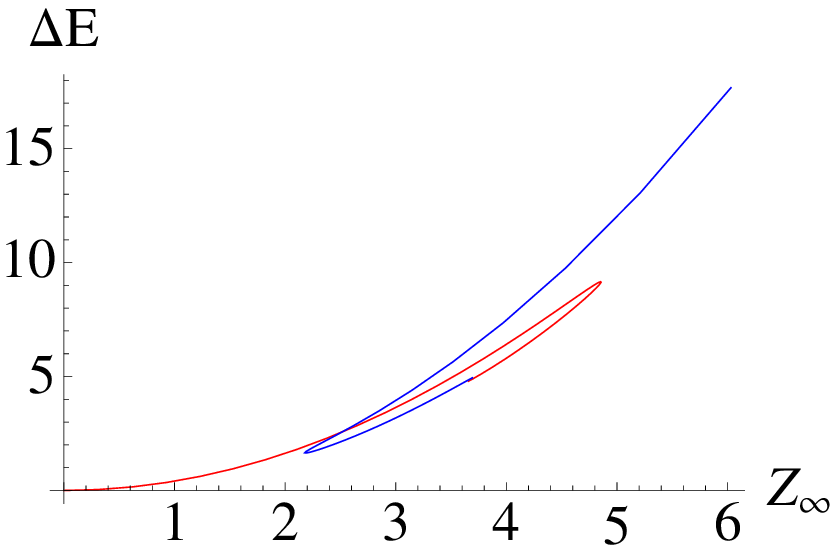} 
\noindent\hfil\includegraphics[scale=.5]{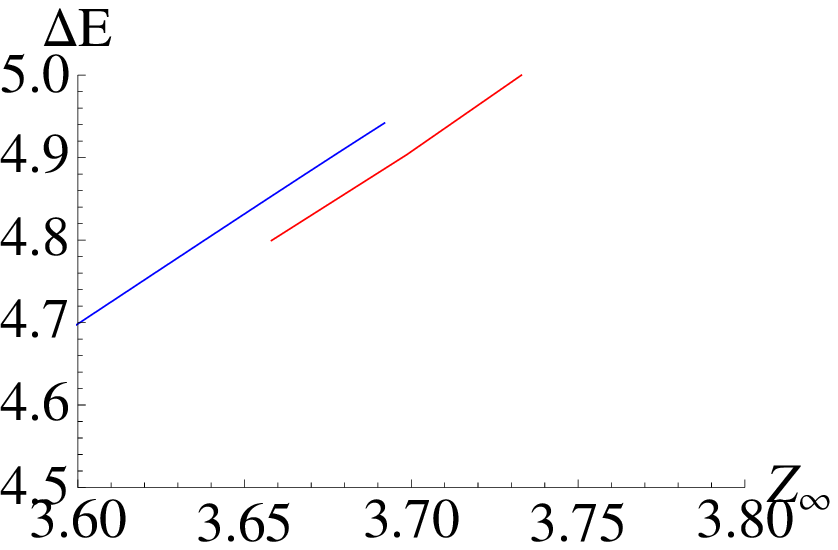} 
\noindent\hfil\includegraphics[scale=.5]{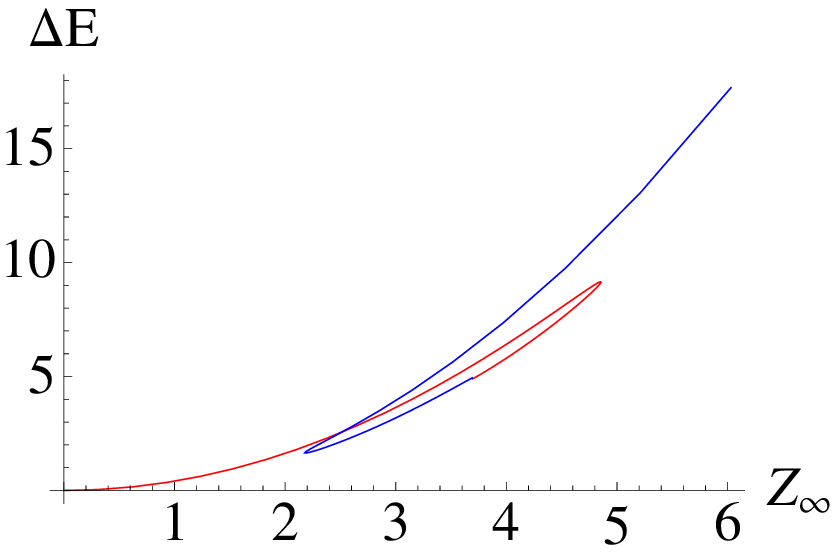} 
\noindent\hfil\includegraphics[scale=.5]{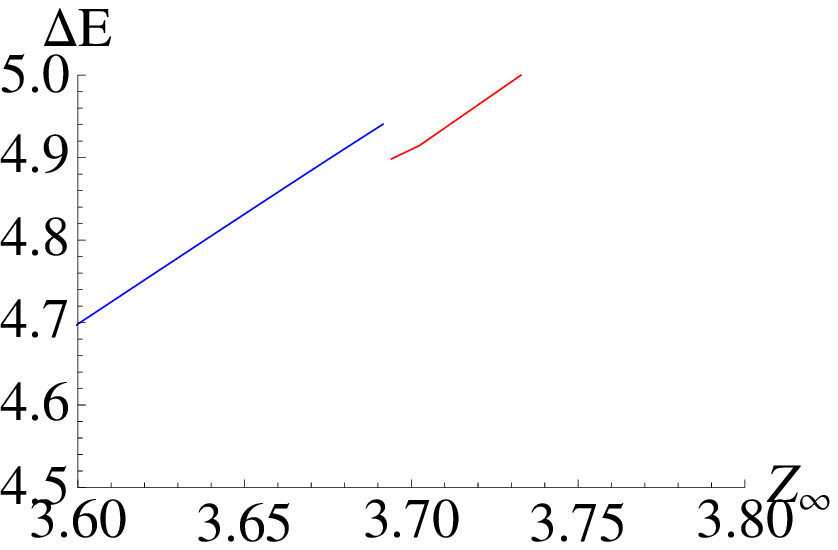}
\caption{The pictures on this figure are the corresponding ones to FIG.~7, in the case of $N=6$ bulk dimensions.}
\end{figure}

For the discussion of the properties of this possible transition we use the method presented 
by Flachi {\it et.~al} in \cite{Flachi}, where a quasi-static evolution of a brane has been considered from the equatorial configuration. In their work, the authors demonstrated that if one plots the difference in the energy $\Delta E$ between a brane configuration (that is quasi-statically evolved from the equatorial configuration and uniquely defined with its initial parameter $g=\theta_0$ or $r_1$) and the equatorial brane configuration, with respect to the cylindrical distance parameter
\[
Z_{\infty}(g)=r_{\infty}\cos\theta(g,r_{\infty}) ,
\]
then, the obtained plot exhibits a loop, i.e.~an instability zone, that is usually a typical sign of a first order phase transition in dynamical systems. The parameter $r_{\infty}$ is a practically chosen maximum radial distance value, until the equations are integrated numerically from the horizon. It is so large that it has the property that the near horizon behavior of the function $\Delta E(Z_{\infty})$ is essentially not affected by any further increase in $r_{\infty}$. 

The energy of a brane with a given initial parameter $g$ can be calculated as
\[
E(g)= -\int_{r_0}^{r_{\infty}}\mathcal{L}(g,r)\ dr\ ,
\]
and  
\[
\Delta E(g)= E(g)-E(\tfrac{\pi}{2})\ .
\]
In FIG.~7 and FIG.~8 we have plotted the curves of $\Delta E(g)[Z_{\infty}(g)]$ with varying $g\equiv\theta_0$ in the region of $(\frac{\pi}{2},\theta_{min})$ for the black hole embedding branch (red curves), and with varying $g\equiv r_1$ in the region of $(1.001,10)$ for the Minkowski embedding branch (blue curves), in $N=5$ and $6$ bulk dimensions respectively. 
In both figures, the top left picture represents the unperturbed, thin brane solutions, and the 
bottom left picture the thick ones. In the plots it is easy to see the presence of the loops which are the sign of a first order phase transition. This implies that small thickness perturbations induced by curvature corrections in the effective brane action do not change the qualitative behavior of the transition in the dimension where a regular perturbative solution exists.

To make the effects of the corrections visible, we enlarged the region where the 
perturbations are getting larger, i.e.~near the singularity. We zoomed into the close neighborhood of the initial parameter $\theta_{min}$ (listed in TABLE.~I) for both the thin and thick curves. The results can be seen on the right hand side pictures of both figures. As a conclusion we can say that within the perturbative approach, small thickness perturbations induced by curvature corrections in the effective brane action have small effects on the $\Delta E(Z_{\infty})$ 
plots of the system, and they do not change the qualitative behavior of the phase transition.

It is very important to note however, that the obtained results are valid only for the exceptional cases where the problem can be regularly solved within the perturbative approach. 
In principle, one can expect a different solution to the thick brane equation with a non-perturbative treatment in the subcritical cases, thus the behavior of the phase transition can also be expected to be different from the picture above.

\section{Conclusions}

In this paper we studied thickness perturbations to static, $D$-dimensional, Dirac-Nambu-Goto branes embedded into higher dimensional, spherically symmetric, black hole spacetimes. The perturbations originate from higher order, curvature corrections added to the thin brane action \cite{Carter,Carter2}, and are quadratic in the thickness of the brane. Following the treatment of \cite{Carter2}, and applying a linear perturbation method with the perturbation parameter $\ve=\ell^2/L^2$, we derived the general form of the perturbation equation for a brane
that is axisymmetric and has a form of a $(D-1)$-dimensional plane at asymptotic infinity.

From the analysis of the asymptotic behavior of the perturbation equation, we found that 
there is no regular solution of the perturbed problem in the Minkowski embedding case, unless the brane is a string, or a sheet. This restriction, however does not hold for the black hole embedding solutions, which are always regular within our perturbative approach. 

From the above results and also from the similar findings of \cite{FG}, we concluded that the absence of regular solutions above the dimension $D=3$ implies, that the problem can not be solved within perturbative approaches around the thin solution which is not smooth on the axis of the system. Hence, for a general discussion, one needs to find a new, exact solution of the curvature corrected problem, that is expected to behave differently from the thin solution with being smooth on the axis. Finding this new solution however seems to be a difficult task, as the equation of motion is a very complicated and highly nonlinear one. 
 
After the above conclusions, we provided the solution of the perturbation equation for various brane ($D$) and bulk ($N$) dimensions. The far distance equations are integrated analytically, while the near horizon solutions are obtained by numerical computations. The deformations of the perturbed brane configurations are plotted and a comparison is made with the corresponding thin brane configurations with identical boundary conditions, for both types of solution.

One motivation of this paper was to consider the effects of higher derivative, curvature corrections on the first order phase transition between the Minkowski and black hole branch, 
that is present in the unperturbed system \cite{MMT1,MMT2,Flachi}. With the solution of the perturbed problem we found that within a perturbative approach, one can consider a phase transition between the two branches only in the cases of $D=2$ or $3$. We investigated the properties of this transition in the case of a $D=3$ brane, and found that small thickness perturbations do not modify the qualitative behavior of the phase transition, i.e.~it remains a first order one, just as in the case of zero thickness.

Since our perturbative approach does not provide a regular thick brane  solution for dimensions $D>3$, we cannot answer, in the most general way, the question whether higher order, curvature corrections in the effective brane action can change the order of the phase transition in the BBH system or not. Although we expect that small corrections may not change the picture too much, as they are quadratic in the thickness of the brane, however a definitive answer can only be given after a new, non-perturbative solution has been found. Further study to address this question is in progress and we hope to report on it in a forthcoming publication.

\begin{acknowledgements}
Some parts of the calculations were performed and checked using the computer algebra programs 
MAPLE, GRTensorII and MATHEMATICA. The authors would like to thank Takahiro Tanaka and Misao Sasaki for helpful discussions on the topic. V.G.Cz.~wishes to express his special thanks to Stefano Ansoldi for his comments on improving the manuscript, and to \'Arp\'ad Luk\'acs for his continuous help and interest during the completion of the work, especially in the numerical part. The research was supported in part by the Japan Society for the Promotion of Science, Contract Nos.~P06816, 19GS0219, 20740133, and the Hungarian National Research Fund, OTKA No.~K67790 grant.
\end{acknowledgements}

\appendix*
\section{Coefficient Functions}
\begin{widetext}
\begin{eqnarray}
A_0&=&-64(a+b)r^4f^2F^4,\\
A_1&=&-64(a+b)fFr^3\csc\theta(2\dot\theta fFnr\cos\theta+(4\dot fr+f(2(2+n)
+\dot\theta r^2(\dot\theta(2f(n-3)-10fr\ddot\theta-\dot fr))))\sin\theta),\\
A_2&=&-160(a+b)f^3r^6(6\dot\theta^2fr^2-1),\\
A_3&=&48(a+b)f^2r^4\csc\theta(\dot\theta r(6\dot\theta fnr\cos\theta 
+8\dot\theta^3f^2nr^3\cos\theta-5\dot\theta^2fr^2(\dot fr-2f(n-3))\sin\theta\nonumber\\ 
&+& 10(f(4+n)+3\dot fr)\sin\theta)-2n\cos\theta),\\
A_4&=&-16\dot\theta fF^2r^4(2a+b+a\dot\theta^2fr^2),
\end{eqnarray}
\begin{eqnarray}
A_5&=&4Fr^2\csc\theta(4an\cos\theta+r(-2\dot\theta^2fnr(7a+4b
+\dot\theta^2fr^2(13a+4b+4a\dot\theta^2fr^2))\cos\theta\nonumber\\ 
&+&(2a\dot\theta^7f^4(1+n)r^6-8(2a+b)\dot f \dot\theta r+\dot\theta^4f^3r^4(\dot\theta(8a-6b-6an)+8a\ddot\theta r+(a+b)\dot f\dot\theta^3r^3)
\nonumber\\
&+&4\dot\theta^2f^2r^2(\dot\theta(2a+b-2(4a+b)n)
+(7a+6b)\ddot\theta r+(2a+b)\dot f\dot\theta^3r^3)+4f(-2\dot\theta(3a(2+n)+b(5+n))\nonumber\\
&-&(10a+9b)\ddot\theta r+(4a+5b)\dot f\dot\theta^3r^3))\sin\theta)),
\end{eqnarray}
\begin{eqnarray}
A_6&=&2 \dot\theta^4r^5(8 \ddot\theta (-71(a+b)+45an+42bn+2(a-2b)n^2)r+ 
4 \dot\theta (b(-80+(69-7n)n)\nonumber\\
&+&a(-80+n(54+(29-3n)n))+2 \ddot\theta (a(37-17n)+b(25-8n))nr^2\cot\theta) 
-4 \dot\theta^4nr^3\cot\theta(2n(-b+an)\nonumber\\
&+&3 \dot f (a+2b+3an)r+2(-2+n)(2a+b-an)\csc^2\theta)
+ \dot f \dot\theta^5r^5((3a+b) \dot f (1+n)r\nonumber\\
&+&4an(n-(-1+n)\csc^2\theta))+4 \dot\theta^3r^2(\dot f (2b(8+n)
+a(13+n(27+5n)))r+n(-a(-25+n)n\nonumber\\
&+&b(-3+11n)+(b(13-11n)+a(28+(-27+n)n))\csc^2\theta)) 
+4 \dot\theta^2 r(n(b(57-31n)\nonumber\\
&+&a(89-n(26+19n)))\cot\theta+ \ddot\theta r^2(3 \dot f (a+2b+3an)r 
+2n(-b+5an+4bn\nonumber\\
&+&(6a+5b-5an-4bn)\csc^2\theta)))),
\end{eqnarray}
\begin{eqnarray}
A_7&=&\dot\theta^2 r^3 (-8 \ddot\theta (a (-240 + n (-68 + 13 n)) + 
      b (-240 + n (-47 + 16 n))) r - (a + b) \dot f^3 \dot\theta^7 r^9 \nonumber\\
&+& 2 (5 a + b) \dot f^2 \dot\theta^6 n r^7 \cot\theta + 
   8 \dot\theta (b (48 - 13 (-4 + n) n) + a (48 + n (60 + (8 - 5 n) n)) \nonumber\\
&-& 2 \ddot\theta n (16 b n + a (-9 + 34 n)) r^2 \cot\theta) + 
   8 \dot\theta^2 r (n (b (42 - 43 n) + a (78 - n (76 + 15 n))) \cot\theta \nonumber\\
&+& \ddot\theta r^2 (16 (a + b) n + \dot f (-14 a - 23 b + 43 a n + 16 b n) r + 
         n (16 a + 19 b - 13 a n - 16 b n) \cot\theta^2)) \nonumber\\
&+& 2 \dot\theta^4 r^3 (-(5 a + b) \ddot\theta \dot f^2 r^4 + 
      4 n \cot\theta (n (7 b - 5 a n) - 
         \dot f (7 a + 14 b + 42 a n + 4 b n) r \nonumber\\
&+& (n-2) (-12 a - 7 b + 
            5 a n) \csc^2\theta)) + 
   2 \dot f \dot\theta^5 r^5 (\dot f (5 b (2 + n) + a (14 + 25 n)) r + 
      2 n (-b + 11 a n + 4 b n \nonumber\\
&+& (12 a + 5 b - 11 a n - 4 b n) \csc^2\theta)) + 
   4 \dot\theta^3 r^2 (\dot f r (a (76 + (150 - 19 n) n) + b (58 + (9 - 4 n) n) \nonumber\\
&+& 6 (3 a + 4 b) \ddot\theta n r^2 \cot\theta) + 
      2 n (a (44 - 15 n) n + b (-6 + 17 n) + (b (20 - 17 n) \nonumber\\
&+& a (50 + n (-62 + 15 n))) \csc^2\theta))),
\end{eqnarray}
\begin{eqnarray}
A_8&=& -2 r (8 \ddot\theta (a (2 + n) (4 + 5 n) + b (8 + n (15 + 4 n))) r + 
3 (2 a + b) \dot f^3 \dot\theta^7 r^9 - 23 (a + b) \dot f^2 \dot\theta^6 n r^7 \cot\theta \nonumber\\
&+& 8 \dot\theta n (n (3 b + a (2 + n)) + 
      \ddot\theta (28 a + 25 b + 17 a n + 8 b n) r^2 \cot\theta) + 
   2 \dot f \dot\theta^5 r^5 (-3 \dot f (5 b (-1 + n) + a (4 + 11 n)) r \nonumber\\
&+& (a + 4 b) n (-2 + n + n \cos[2\theta]) \csc^2\theta) + 
   2 \dot\theta^4 r^3 (3 (7 a + 10 b) \ddot\theta \dot f^2 r^4 + 
      2 n \cot\theta (3 n (-3 b + a n) \nonumber\\
&+& \dot f (74 a + 47 b + 53 a n + 8 b n) r + 
         3 (-2 + n) (4 a + 3 b - a n) \csc^2\theta)) + 
   4 \dot\theta^3 r^2 (\dot f r (2 (b (-78 + n (13 + 2 n)) \nonumber\\
&+& 2 a (-42 + n (-5 + 6 n))) + (47 a + 26 b) \ddot\theta n r^2 \cot\theta) + n (b (3 - 9 n) 
+ a n (-21 + 19 n) \nonumber\\
&-& (n-1) (-9 b + a (-24 + 19 n)) \csc^2\theta)) +4 \dot\theta^2 r (n (21 b n + a (-6 + n (50 + n))) \cot\theta \nonumber\\
&+& 2 \ddot\theta r^2 (3 \dot f (a (-48 + n) + b (-46 + 3 n)) r - (a - 
            2 b) n (-2 + n + n \cos[2\theta]) \csc^2\theta))),
\end{eqnarray}
\begin{eqnarray}
A_9&=&2 (-8 (b - a (-2 + n)) (-1 + n) n \cot\theta + 
   r (-9 (2 a + 3 b) \dot f^3 \dot\theta^5 r^7 + 
      2 (-17 a + 4 b) \dot f^2 \dot\theta^4 n r^5 \cot\theta \nonumber\\
&+& 4 \dot\theta (2 \dot f r (-2 a (2 + n) (2 + 3 n) - 
            b (8 + n (15 + n)) - (28 a + 19 b) \ddot\theta n r^2 \cot\theta) \nonumber\\
&+&  n^2 (-(b - 7 a n) \cot\theta^2 - 8 a \csc^2\theta)) - 
      4 \ddot\theta r (2 \dot f (20 a (2 + n) + b (40 + 17 n)) r + 
         n ((b - 7 a n) \cot\theta^2 + 8 a \csc^2\theta)) \nonumber\\
&+& 2 \dot\theta^2 r (3 (46 a + 43 b) \ddot\theta \dot f^2 r^4 + 
         2 n \cot\theta (n (5 b + a n) - \dot f (74 a + 41 b + 4 (4 a + b) n) r\nonumber\\
&-& (n-2) (5 b + a (4 + n)) \csc^2\theta)) + 
      2 \dot f \dot\theta^3 r^3 (\dot f (2 a (62 + n) + b (102 + 7 n)) r + 
         n (3 b - 21 a n + 4 b n \nonumber\\
&+& (b - 4 b n + 3 a (-8 + 7 n)) \csc^2\theta)))),
\end{eqnarray}
\begin{eqnarray}
A_{10}&=&4 (2 (10 a + 7 b) \dot f^3 \dot\theta^3 r^6 + 
   \dot f^2 r^3 (-2 \dot\theta (10 a (2 + n) + 3 b (6 + n)) - 
      2 (18 a + 17 b) \ddot\theta r - (26 a + 7 b) \dot\theta^2 n r \cot\theta) \nonumber\\
&-& n (b + a n) (-4 + n + n \cos[2\theta]) \cot\theta \csc^2\theta + 
   \dot f r (\dot\theta^9 f^5 (1 + n) (3 (a + b) + 5 a n) r^9 - 
      4 (b - 2 a (-1 + n)) n \cot\theta \nonumber\\
&-&  2 \dot\theta n r ((b - 7 a n) \cot\theta^2 + 8 a \csc^2\theta)) + 
   \dot\theta^6 f^5 (a + b + a n) r^7 (2 \ddot\theta (-13 + 7 n) r + 
      2 \dot\theta (-24 + n (17 + n) \nonumber\\
&+& \dot\theta r ((13 - 7 n) n \cot\theta + 
            \dot\theta r (f (-1 + n^2) + 2 n (n - (-1 + n) \csc^2\theta)))))).
\end{eqnarray}
\begin{eqnarray}
\alpha=
&-&\frac{1}{2 r^4} \left[\right.-2 (10 a + 7 b) r^6 \dot f^3 \dot\theta^3 + 2n\cot\theta 
((b+an)((n-2)\csc^2\theta-n)-2ar^2\ddot f) + 2r\dot f(2n(2a+b-2an)\cot\theta \nonumber\\
&+& r\dot\theta(n((b-7an)\cot^2\theta + 8a\csc^2\theta) + 4(2a+b)r^2\ddot f)) 
+ r^3\dot f^2(\dot\theta (20a(2+n)+ 6b(6+n)+(26a+7b)nr\cot\theta\dot\theta)\nonumber\\
&+& 2(18a+17b)r\ddot\theta))\left.\right]
\end{eqnarray}
\begin{eqnarray}
c_1&=&(1/(960 \eta^5 f^2 r_1^6))(3840 (b (180 + n (109 + 8 n)) + a (180 + n (100 + 17 n))) 
+ \eta r_1 (15 \dot f \eta r_1 (64 (b (272 \nonumber\\
&+& n (164 + 13 n)) + a (272 + n (166 + n (34 + n)))) 
+ 16 \dot f \eta^2 (a (39 + 5 n (4 + n)) + b (69 + n (39 + 4 n))) r_1^2 \nonumber\\
&-& 4 \dot f^2 \eta^4 (b (4 + n) + a (14 + 5 n)) r_1^4 + (a + b) \dot f^3 \eta^6 r_1^6) 
+ 120 \eta^6 f^4 (a + b + a n) (-1 + n^2) r_1^3 (-\eta + 9 r_1 \sigma) \nonumber\\
&+& 4 \eta^3 f^3 r_1^2 (-4 (a + b + a n) (-15 \eta^2 (-68 + n (23 + 9 n)) 
+ 5 \eta^4 (14 - 5 n) n r_1 + \eta^6 (-1 + n) n r_1^2 \nonumber\\
&+& 15 \eta (39 + n (-79 + 28 n)) r_1 \sigma - 45 \eta^3 n (1 + 2 n) r_1^2 \sigma 
+ 180 (-1 + n) n r_1^2 \sigma^2) \nonumber\\
&+& 15 \eta^3 (1 + n) r_1^2 (\dot f (4 \eta (-1 + n) (a + b + a n) 
+ 9 (3 (a + b) + 5 a n) r_1 \sigma) + r_1 (2 a f^{(3)} \eta r_1 \nonumber\\
&+& \ddot f (3 b \eta + a (5 \eta (1 + n) + 18 r_1 \sigma))))) 
+ 5 f (64 (b (6 \eta (436 + 3 n (55 + 2 n)) + \eta^3 n (62 + 19 n) r_1 \nonumber\\
&+& 3 (2652 + n (1397 + 91 n)) r_1 \sigma) + a (6 \eta (436 + n (190 + n (15 + 2 n))) 
+ \eta^3 n (8 + n) (10 + 3 n) r_1\nonumber\\
&+& 3 (2652 + n (1304 + n (181 + 3 n))) r_1 \sigma)) - \eta^2 r_1^2 (48 \ddot f \eta (b (60 + n (35 + 4 n)) + a (36 + n (32 + 5 n))) r_1 \nonumber\\
&+& 3 (a + b) \dot f^3 \eta^4 r_1^4 (2 \eta + 9 r_1 \sigma) 
+ \dot f^2 \eta^2 r_1^2 (9 (a + b) \ddot f \eta^3 r_1^3 
+ 4 (-6 \eta (5 a + 7 b + 2 (8 a + b) n) + (5 a + b) \eta^3 n r_1 \nonumber\\
&-& 3 (5 a + b) (-3 + 5 n) r_1 \sigma)) - 8 \dot f (3 \ddot f \eta^3 (b (5 + n) 
+ a (17 + 5 n)) r_1^3 + 2 (a (-6 \eta (-58 + n (16 + 25 n)) \nonumber\\
&+& 2 \eta^3 n (12 + 5 n) r_1 + 3 (564 + (132 - 5 n) n) r_1 \sigma) 
+ b (-6 \eta (-94 + n (5 + 4 n)) + 2 \eta^3 n (13 + 4 n) r_1 \nonumber\\ 
&+& 3 (564 + (173 - 4 n) n) r_1 \sigma))))) 
+ 2 \eta f^2 r_1 (16 (b (-30 \eta^2 (-157 + n (3 + 14 n)) 
+ 20 \eta^4 n (13 + n) r_1 - 7 \eta^6 n r_1^2 \nonumber\\
&+& 15 \eta (1884 + (115 - 71 n) n) r_1 \sigma + 5 \eta^3 n (66 + 13 n) r_1^2 \sigma 
+ 45 (1392 + n (527 + 13 n)) r_1^2 \sigma^2) \nonumber\\
&+& a (-30 \eta^2 (-157 + n + 36 n^2 + 7 n^3) 
+ 20 \eta^4 n (27 - 2 (-2 + n) n) r_1 + \eta^6 n (-10 + n (9 + 4 n)) r_1^2 \nonumber\\
&+& 15 \eta (1884 + n (276 + (-191 + n) n)) r_1 \sigma 
- 75 \eta^3 n (-14 + (-2 + n) n) r_1^2 \sigma \nonumber\\
&+& 45 (1392 + n (522 - (-19 + n) n)) r_1^2 \sigma^2)) 
+ \eta^2 r_1^2 (15 \dot f^2 \eta^3 (1 + n) r_1^2 (\eta (7 b + a (9 + 10 n)) 
+ 9 (3 a + b) r_1 \sigma) \nonumber\\
&-& 10 \eta r_1 (3 a f^{(3)} \eta (32 + n (15 + n)) r_1 
- 3 (a + b) \ddot f^2 \eta^3 r_1^3 + 2 \ddot f (3 b \eta (3 + n) (8 + n) 
+ a (3 \eta (-2 + n (43 + n (18 + n))) \nonumber\\
&-& \eta^3 n (9 + 2 n) r_1 
+ 3 (-24 + n (33 + 7 n)) r_1 \sigma))) + 2 \dot f (15 \eta^3 r_1^3 ((a + b) f^{(3)} \eta r_1 
+ \ddot f (\eta (b (3 + n) + a (7 + 5 n))\nonumber\\
&+& 9 (a + b) r_1 \sigma)) + 2 (10 b \eta (18 \eta (7 + n - n^2) + \eta^3 n (7 + 2 n) r_1 
- 3 (-18 + n (23 + 7 n)) r_1 \sigma) \nonumber\\
&+& a (90 \eta^2 (11 + n (20 + n - 2 n^2)) 
+ 20 \eta^4 n (2 + n (6 + n)) r_1 - \eta^6 (-1 + n) n r_1^2 \nonumber\\
&-& 30 \eta (-9 + n (-19 + n (45 + 7 n))) r_1 \sigma + 45 \eta^3 n (1 + 2 n) r_1^2 \sigma 
- 180 (-1 + n) n r_1^2 \sigma^2)))))))\left.\right|_{r_1}
\end{eqnarray}
\begin{eqnarray}
c_3&=&(1/(192 \eta^3 f r_1^4))(-192 (b (20 + n (47 + 9 n)) + a (20 + n (40 + n (15 + n))))
\nonumber\\ 
&+& \eta r_1 (24 \eta^5 f^3 (a + b + a n) (-1 + n^2) r_1^2 - 3 \dot f \eta r_1 (16 (b (4 + n) (3 + 4 n) + a (12 + n (12 + 5 n)))\nonumber\\ 
&-& 4 (5 a + b) \dot f \eta^2 (1 + n) r_1^2 + (a + b) \dot f^2 \eta^4 r_1^4) + 4 \eta^2 f^2 r_1 (3 \eta^3 (1 + n) r_1^2 (\dot f (3 (a + b) + 5 a n) + 2 a \ddot f r_1)\nonumber\\ 
&+& 4 (a + b + a n) (3 \eta (13 + (5 - 6 n) n) + \eta^3 n (1 + 2 n) r_1 - 21 (-1 + n) n r_1 \sigma)) - 2 f (16 (b (3 \eta (36 - 5 (-5 + n) n)\nonumber\\
&+& \eta^3 n (2 + 5 n) r_1 + 3 (96 + 23 n (7 + n)) r_1 \sigma) + a (-3 \eta (1 + n) (-36 + 5 n^2) + \eta^3 n (6 + n (10 + 3 n)) r_1 \nonumber\\
&+& 3 (96 + n (150 + (37 - 3 n) n)) r_1 \sigma)) + \eta^2 r_1^2 (12 a \ddot f \eta (8 + 
n (7 + n)) r_1 - 3 (3 a + b) \dot f^2 \eta^3 (1 + n) r_1^2\nonumber \\
&+& 2 \dot f (-3 (a + b) \ddot f \eta^3 r_1^3 
+ 2 (6 b \eta (2 + n) (3 + n) + a (6 \eta (1 + n) (3 + n (8 + n)) -\eta^3 n (1 + 2 n) r_1 
\nonumber\\
&+& \left.21 (-1 + n) n r_1 \sigma)))))))\right|_{r_1}
\end{eqnarray}
\begin{eqnarray}
S&=&-\frac{1}{480(a+b)r^4\varepsilon f^2F^2}\left[\right.
n (-8 \varepsilon (1 + 2 n) (b + a n) + 20 r^2 + 
   20 \dot\theta^6 \varepsilon f^4 (-13 + 7 n) (a + b + a n) r^6 
\nonumber\\
&+& 10 \varepsilon r (-4 a \ddot f r + \dot f (8 a + 4 b - 8 a n + (26 a + 7 b) \dot f \dot\theta^2 r^3)) + 4 \dot\theta^2 f^3 r^4 (20 \dot\theta^2 \varepsilon (b (-11 + 6 n) 
\nonumber\\
&+& a (-19 + n (8 + 3 n))) - 240 (a + b) \ddot\theta^2 \varepsilon r^2 + 
      10 \dot\theta \varepsilon r (\ddot\theta (-37 a - 25 b + 17 a n + 8 b n) 
\nonumber\\
&+&  8 (a + b) \theta^{(3)} r) + \dot\theta^4 r^2 (4 \varepsilon (-b (1 + 2 n) + 2 a (-1 + n + n^2)) + 5 r^2 + 5 \varepsilon r (3 \dot f (a + 2 b + 3 a n) 
\nonumber\\
&+& 4 a \ddot f r))) + f^2 r^2 (20 \dot\theta^2 \varepsilon (b (2 + 19 n) + a (-2 + n (44 + 3 n))) + 240 (a + b) \ddot\theta^2 \varepsilon r^2 
\nonumber\\
&-& 60 (3 a + 4 b) \ddot\theta \dot f \dot\theta^3 \varepsilon r^4 - 
      5 (5 a + b) \dot f^2 \dot\theta^6 \varepsilon r^6 + 
      40 \dot\theta \varepsilon r (\ddot\theta (28 a + 25 b + 17 a n + 8 b n) 
\nonumber\\
&+& 8 (a + b) \theta^{(3)} r) +  4 \dot\theta^4 r^2 (2 \varepsilon (-5 (b + 2 b n) + a (-8 + n (7 + 6 n))) +  15 r^2 
\nonumber\\
&+& 5 \varepsilon r (\dot f (4 b (2 + n) + a (4 + 33 n)) + (13 a + 4 b) \ddot f r))) + 
   2 f (20 \varepsilon (b - a (-2 + n)) (-1 + n) 
\nonumber\\
&+& \dot\theta r^2 (20 (28 a + 19 b) \ddot\theta \dot f \varepsilon r^2 + 
         \dot\theta (8 \varepsilon (-2 (a + b) + (a - 4 b) n) + 30 r^2 + 
            5 \varepsilon r (2 (7 a + 4 b) \ddot f r 
\nonumber\\
&+& \dot f (78 b + 8 b n + 20 a (7 + 2 n) - (9 a + 11 b) \dot f \dot\theta^2 r^3))))))
\left.\right]\ ,
\end{eqnarray}

\begin{eqnarray}\label{S1}
S_1&=&\frac{1}{32(a+b)r^4\varepsilon f^2F^2}\left[
n (-4 \varepsilon n (b + a n) + 4 r^2 + 4 \dot\theta^6 \varepsilon f^4 (-13 + 7 n) (a + b + a n) r^6 \right.
\nonumber\\
&+& 2 \varepsilon r (-4 a \ddot f r + \dot f (8 a + 4 b - 8 a n + (26 a + 7 b) \dot f \dot\theta^2 r^3)) + 4 \dot\theta^2 f^3 r^4 (4 \dot\theta^2 \varepsilon (b (-11 + 6 n) 
\nonumber\\
&+& a (-19 + n (8 + 3 n))) - 48 (a + b) \ddot\theta^2 \varepsilon r^2 + 
      2 \dot\theta \varepsilon r (\ddot\theta (-37 a - 25 b + 17 a n + 8 b n) 
\nonumber\\
&+& 8 (a + b) \theta^{(3)} r) + \dot\theta^4 r^2 (2 \varepsilon n (-b + a n) + r^2 + 
         \varepsilon r (3 \dot f (a + 2 b + 3 a n) + 4 a \ddot f r))) 
\nonumber\\
&+& f^2 r^2 (4 \dot\theta^2 \varepsilon (b (2 + 19 n) + a (-2 + n (44 + 3 n))) + 
      48 (a + b) \ddot\theta^2 \varepsilon r^2 
\nonumber\\
&-& 12 (3 a + 4 b) \ddot\theta \dot f \dot\theta^3 \varepsilon r^4 - (5 a + 
         b) \dot f^2 \dot\theta^6 \varepsilon r^6 + 
      8 \dot\theta \varepsilon r (\ddot\theta (28 a + 25 b + 17 a n + 8 b n) 
\nonumber\\
&+&  8 (a + b) \theta^{(3)} r) +  4 \dot\theta^4 r^2 (\varepsilon n (-5 b + 3 a n) + 3 r^2 + 
 \varepsilon r (\dot f (4 b (2 + n) + a (4 + 33 n)) 
\nonumber\\
&+& (13 a + 4 b) \ddot f r))) + 
   2 f (4 \varepsilon (b - a (-2 + n)) (-1 + n) + 
      \dot\theta r^2 (4 (28 a + 19 b) \ddot\theta \dot f \varepsilon r^2 
\nonumber\\
&+& \dot\theta (-8 b \varepsilon n + 6 r^2 + \varepsilon r (2 (7 a + 4 b) \ddot f r + 
\dot f (78 b + 8 b n + 20 a (7 + 2 n) 
\nonumber\\
&-& (9 a + 11 b) \dot f \dot\theta^2 r^3)))))) \left.\right]\ ,
\end{eqnarray}
\begin{eqnarray}\label{S2}
S_2&=&\frac{1}{16(a+b)r^3f^2}\left[
n(2\dot f\dot\theta(b+a(8-7n))r+4\dot\theta^5f^3(n-1)(a+b+an)r^4\right. 
\nonumber\\
&+&f(2\dot\theta n(8a+b-7an)+2\ddot\theta(8a+b-7an)r+\dot f\dot\theta^3(7an+4bn-8a-5b)r^3) 
\nonumber\\
&+&
2\dot\theta^2f^2r^2(2\ddot\theta(b(4n-5)+a(5n-6))r+\dot\theta(b(7n-9)+a((27-5n)n-24) 
\nonumber\\
&+&\left. a\dot f\dot\theta^2(n-1)r^3)))\right]\ ,
\end{eqnarray}

\begin{equation}\label{S3}
S_3=\frac{1}{8(a+b)r^4f^2}\left[n(n-2)(1+r^2f\dot\theta^2)(b+an+2(b-a(n-2))r^2f\dot\theta^2) \right]\ ,
\end{equation}

\newpage
{\small
\begin{eqnarray}\label{T2}
T_2&=&\frac{1}{64(a+b)\varepsilon f^2r^4F^4}\left[\right.-8 \dot\theta^9 \varepsilon f^6 (a + b + a n) (-1 + n^2) r^9 
\nonumber\\
&-& 2 \dot\theta^2 f^4 r^5 (240 (a + b) \ddot\theta^2 \dot\theta \varepsilon (-3 + n) r^2 - 
480 (a + b) \ddot\theta^3 \varepsilon r^3 + 8 \ddot\theta \dot\theta^2 \varepsilon r (-71 (a + b) \nonumber\\
&+& 45 a n + 42 b n + 2 (a - 2 b) n^2 + 24 (a + b) \ddot\theta n r^2 \cot\theta) 
+ 4 \dot\theta^3 \varepsilon (b (-80 + (69 - 7 n) n) \nonumber\\
&+& a (-80 + n (54 + (29 - 3 n) n)) + 2 \ddot\theta (a (37 - 17 n) + b (25 - 8 n)) n r^2 \cot\theta)
\nonumber\\
&+&  4 \dot\theta^6 n r^3 \cot\theta (2 \varepsilon n (b - a n) - r^2 + 
\varepsilon r (-3 \dot f (a + 2 b + 3 a n) - 4 a \ddot f r) 
\nonumber\\
&-& 2 \varepsilon (b - a (-2 + n)) (-2 + n) \csc\theta^2) 
+ 4 \dot\theta^5 r^2 (\varepsilon n (-a (-25 + n) n + b (-3 + 11 n)) 
\nonumber\\
&+& (5 + 4 n) r^2 + \varepsilon r (\dot f (2 b (8 + n) + a (13 + n (27 + 5 n))) + 
r (\ddot f (5 a - 3 b - 2 a n) - 2 a f^{(3)} r)) 
\nonumber\\
&+& \varepsilon n (b (13 - 11 n) + a (28 + (-27 + n) n)) \csc\theta^2) + 
 4 \dot\theta^4 r (\varepsilon n (b (57 - 31 n) 
\nonumber\\
&+& a (89 - n (26 + 19 n))) \cot\theta +\ddot\theta r^2 (2 \varepsilon n (5 a n + b (-1 + 4 n)) + r^2 + \varepsilon r (3 \dot f (a + 2 b + 3 a n) 
\nonumber\\
&+& 4 a \ddot f r) + 2 \varepsilon (a (6 - 5 n) + b (5 - 4 n)) n \csc\theta^2))
+ \dot f \dot\theta^7 r^5 ((3 a + b) \dot f \varepsilon (1 + n) r + 2 (2 a \varepsilon n^2 
\nonumber\\
&+&  r^2 + (a + b) \ddot f \varepsilon r^2 - 2 a \varepsilon (-1 + n) n \csc\theta^2))) 
+  2 f^2 r (3 (2 a + b) \dot f^3 \dot\theta^7 \varepsilon r^9 
\nonumber\\
&-&  23 (a + b) \dot f^2 \dot\theta^6 \varepsilon n r^7 \cot\theta + 
 8 \ddot\theta \varepsilon r (a (2 + n) (4 + 5 n) + b (8 + n (15 + 4 n)) 
\nonumber\\
&+&  6 (a + b) \ddot\theta n r^2 \cot\theta) + 8 \dot\theta \varepsilon (n^2 (3 b + a (2 + n)) - 90 (a + b) \ddot\theta^2 \dot f r^5 + \ddot\theta n (b (25 + 8 n) 
\nonumber\\
&+& a (28 + 17 n)) r^2 \cot\theta) - 2 \dot f \dot\theta^5 r^5 (2 (a + 4 b) \varepsilon n^2 + 9 r^2 + 3 \varepsilon r (\dot f (5 b (-1 + n) + a (4 + 11 n)) 
\nonumber\\
&+& 8 (a + b) \ddot f r) -  2 (a + 4 b) \varepsilon (-1 + n) n \csc\theta^2) 
+ 4 \dot\theta^3 r^2 (\varepsilon n (b (3 - 9 n) + a n (-21 + 19 n)) 
\nonumber\\
&+& (-7 - 4 n) r^2 + \varepsilon r (2 r (\ddot f (b (9 + 4 n) + 2 a (5 + 7 n)) 
+ (5 a + 2 b) f^{(3)} r) + \dot f (2 (b (-78 
\nonumber\\
&+&  n (13 + 2 n)) + 2 a (-42 + n (-5 + 6 n))) + (47 a + 26 b) \ddot\theta n r^2 \cot\theta)) 
- \varepsilon (-1 + n) n (-9 b 
\nonumber\\
&+& a (-24 + 19 n)) \csc\theta^2) + 2 \dot\theta^4 r^3 (3 (7 a + 10 b) \ddot\theta \dot f^2 \varepsilon r^4 + 2 \varepsilon n r (\dot f (b (47 + 8 n) + a (74 + 53 n)) 
\nonumber\\
&+&  4 (5 a + 2 b) \ddot f r) \cot\theta + 6 n \cot\theta (\varepsilon n (-3 b + a n) 
+ 2 r^2 + \varepsilon (3 b - a (-4 + n)) (-2 + n) \csc\theta^2)) 
\nonumber\\
&+&  4 \dot\theta^2 r (\varepsilon n (21 b n + a (-6 + n (50 + n))) \cot\theta + 
 \ddot\theta r^2 (4 (a - 2 b) \varepsilon n^2 - 3 r^2 + 6 \varepsilon r (\dot f (a (-48 + n) 
\nonumber\\
&+& b (-46 + 3 n)) + (a + b) \ddot f r) - 4 (a - 2 b) \varepsilon (-1 + n) n \csc\theta^2))) - 
  4 \dot\theta^6 f^5 r^7 (2 \dot\theta \varepsilon (a + b 
\nonumber\\
&+&  a n) (-24 + n (17 + n)) + 2 \ddot\theta \varepsilon (-13 + 7 n) (a + b + a n) r 
- 2 \dot\theta^2 \varepsilon n (-13 + 7 n) (a + b 
\nonumber\\
&+& a n) r \cot\theta +  \dot\theta^3 r^2 (\varepsilon (1 + n) r (\dot f (3 (a + b) + 5 a n) + 2 a \ddot f r) + 2 (2 \varepsilon n^2 (a + b + a n) 
\nonumber\\
&+&(1 + n) r^2 - 2 \varepsilon (-1 + n) n (a + b + a n) \csc\theta^2))) 
-  2 f (-9 (2 a + 3 b) \dot f^3 \dot\theta^5 \varepsilon r^8 
\nonumber\\
&-& 8 \varepsilon (b - a (-2 + n)) (-1 + n) n \cot\theta +  2 (-17 a + 4 b) \dot f^2 \dot\theta^4 \varepsilon n r^6 \cot\theta 
- 4 \ddot\theta r^2 (-r^2 
\nonumber\\
&+& 2 \varepsilon r (\dot f (20 a (2 + n) + b (40 + 17 n)) + (10 a + 9 b) \ddot f r) + \varepsilon n (b - 7 a n) \cot\theta^2 + 8 a \varepsilon n \csc\theta^2) 
\nonumber\\
&-& 4 \dot\theta r (\varepsilon n^2 (-b + 7 a n) + (-2 - n) r^2 
+ 2 \varepsilon r (r (2 \ddot f (3 a (2 + n) + b (5 + n)) + (2 a + b) f^{(3)} r) 
\nonumber\\
&+& \dot f (2 a (2 + n) (2 + 3 n) +  b (8 + n (15 + n)) + (28 a + 19 b) \ddot\theta n r^2 \cot\theta)) + \varepsilon (b 
\nonumber\\
&+&  a (8 - 7 n)) n^2 \csc\theta^2) + 2 \dot f \dot\theta^3 r^4 (\varepsilon n (-21 a n + b (3 + 4 n)) + 7 r^2 + \varepsilon r (\dot f (2 a (62 + n) 
\nonumber\\
&+&  b (102 + 7 n)) + 12 b \ddot f r)+ \varepsilon n (b - 4 b n + 3 a (-8 + 7 n)) \csc\theta^2) -  2 \dot\theta^2 r^2 (-3 (46 a + 43 b) \ddot\theta \dot f^2 \varepsilon r^4 
\nonumber\\
&+& 2 n \cot\theta (-\varepsilon n (5 b + a n) + 4 r^2 + \varepsilon r (\dot f (74 a + 41 b + 4 (4 a + b) n) + (5 a + 4 b) \ddot f r) 
\nonumber\\
&+& \varepsilon (-2 + n) (5 b + a (4 + n)) \csc\theta^2))) + 
  f^3 r^3 (-480 (a + b) \ddot\theta^2 \dot\theta \varepsilon (4 + n) r^2 
\nonumber\\
&-& 160 (a + b) \ddot\theta^3 \varepsilon r^3 + (a + b) \dot f^3 \dot\theta^9 \varepsilon r^9 - 
     2 (5 a + b) \dot f^2 \dot\theta^8 \varepsilon n r^7 \cot\theta 
\nonumber\\
&-&  8 \ddot\theta \dot\theta^2 \varepsilon r (b (240 + (47 - 16 n) n) + 
        a (240 + (68 - 13 n) n) + 36 (a + b) \ddot\theta n r^2 \cot\theta) 
\nonumber\\
&+&  8 \dot\theta^3 \varepsilon (b (-48 + 13 (-4 + n) n) + 
        a (2 + n) (-24 + n (-18 + 5 n)) + 30 (a + b) \ddot\theta^2 \dot f r^5 
\nonumber\\
&+&  2 \ddot\theta n (16 b n + a (-9 + 34 n)) r^2 \cot\theta) + 
     8 \dot\theta^4 r (\varepsilon n (b (-42 + 43 n) + a (-78 + n (76 + 15 n))) \cot\theta 
\nonumber\\
&-& \ddot\theta r^2 (16 (a + b) \varepsilon n + 3 r^2 + \varepsilon r (\dot f (-14 a - 23 b + 43 a n + 16 b n) + 6 (3 a + 2 b) \ddot f r) 
\nonumber\\
&+&  \varepsilon (b (19 - 16 n) + a (16 - 13 n)) n \cot\theta^2)) + 
     2 \dot\theta^6 r^3 ((5 a + b) \ddot\theta \dot f^2 \varepsilon r^4 + 
        4 n \cot\theta (-7 b \varepsilon n + 5 a \varepsilon n^2 
\nonumber\\
&+&  4 r^2 + \varepsilon r (\dot f (2 b (7 + 2 n) + 7 a (1 + 6 n)) + (17 a + 
     4 b) \ddot f r) + \varepsilon (7 b + a (12 - 5 n)) (-2 + n) \csc\theta^2))
\nonumber\\
&+&  2 \dot f \dot\theta^7 r^5 (\varepsilon r (\dot f (-5 b (2 + n) - a (14 + 25 n)) - 
   2 (9 a + 5 b) \ddot f r) + 2 (\varepsilon n (b - 11 a n - 4 b n) - 5 r^2 
\nonumber\\
&+&  \varepsilon n (b (-5 + 4 n) + a (-12 + 11 n)) \csc\theta^2)) - 
     4 \dot\theta^5 r^2 (\varepsilon r (-2 \ddot f (-8 a + b + 19 a n + 4 b n) r 
\nonumber\\
&-&  4 (4 a + b) f^{(3)} r^2 +  \dot f (a (76 + (150 - 19 n) n) + b (58 + (9 - 4 n) n) + 
    6 (3 a + 4 b) \ddot\theta n r^2 \cot\theta)) 
\nonumber\\
&+&  2 (\varepsilon n (a (44 - 15 n) n + b (-6 + 17 n)) + (9 + 6 n) r^2 + \varepsilon n (b (20 - 17 n) 
\nonumber\\
&+&  a (50 + n (-62 + 15 n))) \csc\theta^2))) +  4 (-2 (10 a + 7 b) \dot f^3 \dot\theta^3 \varepsilon r^6 + \dot f^2 \varepsilon r^3 (2 (18 a + 17 b) \ddot\theta r 
\nonumber\\
&+&  \dot\theta (20 a (2 + n) +   6 b (6 + n) + (26 a + 7 b) \dot\theta n r \cot\theta)) + 
     2 n \cot\theta (-\varepsilon n (b + a n) + r^2 - 2 a \ddot f \varepsilon r^2 
\nonumber\\
&+&  \varepsilon (-2 + n) (b + a n) \csc\theta^2) + 
     2 \dot f r (2 \varepsilon n (2 a + b - 2 a n) \cot\theta 
\nonumber\\
&+&  \dot\theta r (-r^2 + 4 (2 a + b) \ddot f \varepsilon r^2 + 
           \varepsilon n ((b - 7 a n) \cot\theta^2 + 8 a \csc\theta^2))))\left.\right]
\end{eqnarray}
}

\begin{eqnarray}
S_0&=&-\frac{1}{192(a+b)r^3\varepsilon f^2F^4}\left[\right.
24 \dot\theta^9 \varepsilon f^6 (a + b + a n) (-1 + n^2) r^8 +4 \dot\theta^6 f^5 r^6 (6 \dot\theta \varepsilon (a +b
\nonumber\\
&+& a n) (-24 + n (17 + n)) + 
    6 \ddot\theta \varepsilon (-13 + 7 n) (a + b + a n) r + 
    \dot\theta^3 r^2 (4 \varepsilon n (1 + 2 n) (a + b + a n) 
\nonumber\\
&+& 6 (1 + n) r^2 + 3 \varepsilon (1 + n) r (\dot f (3 (a + b) + 5 a n) + 2 a \ddot f r))) + 
 2 \dot\theta^2 f^4 r^4 (-12 \dot\theta^3 \varepsilon (a (-1 + n) (-80 
\nonumber\\
&+& n (-26 + 3 n)) +  b (80 + n (-69 + 7 n))) - 1440 (a + b) \ddot\theta^3 \varepsilon r^3 + 
240 (a + b) \ddot\theta \dot\theta \varepsilon r^2 (3 \ddot\theta (-3 + n) 
\nonumber\\
&+& 4 \theta^{(3)} r) + 4 \ddot\theta \dot\theta^4 r^3 (4 \varepsilon n (b + 4 b n + a (3 + 5 n)) + 3 r^2 + 3 \varepsilon r (3 \dot f (a + 2 b + 3 a n) 
\nonumber\\
&+& 4 a \ddot f r)) + \dot f \dot\theta^7 r^5 (4 a \varepsilon n (1 + 2 n) + 6 r^2 + 
       3 \varepsilon r ((3 a + b) \dot f (1 + n) + 2 (a + b) \ddot f r)) 
\nonumber\\
&-& 24 \dot\theta^2 \varepsilon r (\ddot\theta (b (71 - 42 n + 4 n^2) + 
    a (71 - n (45 + 2 n))) + 4 (a + b) r (2 \theta^{(3)} (-3 + n) + \theta^{(4)} r)) 
\nonumber\\
&+& 4 \dot\theta^5 r^2 (2 \varepsilon n (b (2 + 11 n) + a (14 - (-24 + n) n)) + 
       3 (5 + 4 n) r^2 + 3 \varepsilon r (\dot f (2 b (8 + n) 
\nonumber\\
&+& a (13 + n (27 + 5 n)))+ r (\ddot f (5 a - 3 b - 2 a n) - 2 a f^{(3)} r)))) - 2 f (384 (a + b) \theta^{(3)} \dot f \varepsilon r^3 
\nonumber\\
&-& 18 (46 a + 43 b) \ddot\theta \dot f^2 \dot\theta^2 \varepsilon r^5 
+  27 (2 a + 3 b) \dot f^3 \dot\theta^5 \varepsilon r^7 
- 2 \dot f \dot\theta^3 r^3 (2 \varepsilon n (b (5 + 4 n) 
\nonumber\\
&-& 3 a (4 + 7 n)) + 21 r^2 + 3 \varepsilon r (\dot f (2 a (62 + n) + b (102 + 7 n)) + 12 b \ddot f r)) 
\nonumber\\
&+& 4 \ddot\theta r (2 \varepsilon n (-b + a (4 + 7 n)) 
- 3 r^2 + 6 \varepsilon r (\dot f (20 a (2 + n) + b (40 + 17 n)) + (10 a + 9 b) \ddot f r))
\nonumber\\
&+& 4 \dot\theta (2 \varepsilon n^2 (-b + a (4 + 7 n)) - 3 (2 + n) r^2 + 
       6 \varepsilon r (\dot f (2 a (2 + n) (2 + 3 n) + b (8 + n (15 + n))) 
\nonumber\\
&+& r (2 \ddot f (3 a (2 + n) + b (5 + n)) + (2 a + b) f^{(3)} r)))) + 
 f^3 r^2 (480 (a + b) \ddot\theta^3 \varepsilon r^3 
\nonumber\\
&-& 6 (5 a + b) \ddot\theta \dot f^2 \dot\theta^6 \varepsilon r^7 - 3 (a + b) \dot f^3 \dot\theta^9 \varepsilon r^9 + 480 (a + b) \ddot\theta \dot\theta \varepsilon r^2 (3 \ddot\theta (4 + n) 
\nonumber\\
&+& 4 \theta^{(3)} r) + 24 \dot\theta^3 \varepsilon (b (48 - 13 (-4 + n) n) + 
       a (48 + n (60 + (8 - 5 n) n)) - 30 (a + b) \ddot\theta^2 \dot f r^5) 
\nonumber\\
&+& 2 \dot f \dot\theta^7 r^5 (4 \varepsilon n (b + 4 b n + a (6 + 11 n)) + 30 r^2 + 
       3 \varepsilon r (\dot f (5 b (2 + n) + a (14 + 25 n)) 
\nonumber\\
&+& + 2 (9 a + 5 b) \ddot f r)) - 24 \dot\theta^2 \varepsilon r (\ddot\theta (-240 (a + b) - (68 a + 47 b) n + (13 a + 16 b) n^2)
\nonumber\\
&+& 16 (a + b) r (\theta^{(3)} (-1 + 2 n) + \theta^{(4)} r)) +  8 \dot\theta^4 r^3 (24 (a + b) \theta^{(3)} \dot f \varepsilon r^2 + \ddot\theta (2 \varepsilon n (8 a + 5 b + 13 a n 
\nonumber\\
&+& 16 b n) +9 r^2 + 3 \varepsilon r (\dot f (-14 a - 23 b + 43 a n + 16 b n) +  6 (3 a + 2 b) \ddot f r)))
\nonumber\\
&-& 4 \dot\theta^5 r^2 (4 \varepsilon n (-b (1 + 17 n) + 5 a (-5 + n (-7 + 3 n))) - 
       18 (3 + 2 n) r^2 
\nonumber\\
&+& 3 \varepsilon r (\dot f (b (-58 + n (-9 + 4 n)) + a (-76 + n (-150 + 19 n))) + 2 r (\ddot f (-8 a + b + 19 a n + 4 b n) 
\nonumber\\
&+& 2 (4 a + b) f^{(3)} r)))) - 2 f^2 (18 (7 a + 10 b) \ddot\theta \dot f^2 \dot\theta^4 \varepsilon r^7 + 9 (2 a + b) \dot f^3 \dot\theta^7 \varepsilon r^9 
\nonumber\\
&+& 24 \dot\theta \varepsilon (n^2 (3 b + a (2 + n)) - 90 (a + b) \ddot\theta^2 \dot f r^5) - 
    2 \dot f \dot\theta^5 r^5 (2 (a + 4 b) \varepsilon n (1 + 2 n) + 27 r^2 
\nonumber\\
&+& 9 \varepsilon r (\dot f (5 b (-1 + n) + a (4 + 11 n)) + 8 (a + b) \ddot f r)) + 
    24 \varepsilon r (\ddot\theta (a (2 + n) (4 + 5 n) 
\nonumber\\
&+& b (8 + n (15 + 4 n))) + 4 (a + b) r (2 \theta^{(3)} (2 + n) + \theta^{(4)} r)) + 
    4 \dot\theta^2 r^3 (72 (a + b) \theta^{(3)} \dot f \varepsilon r^2 
\nonumber\\
&+& \ddot\theta (4 (a - 2 b) \varepsilon n (1 + 2 n) - 9 r^2 + 
    18 \varepsilon r (\dot f (a (-48 + n) + b (-46 + 3 n)) + (a + b) \ddot f r))) 
\nonumber\\
&+&  4 \dot\theta^3 r^2 (2 \varepsilon n (-9 b n + a (-12 + n (-10 + 19 n))) - 
       3 (7 + 4 n) r^2 + 6 \varepsilon r (\dot f (b (-78 + n (13 + 2 n)) 
\nonumber\\
&+& 2 a (-42 + n (-5 + 6 n))) + r (\ddot f (b (9 + 4 n) + 2 a (5 + 7 n)) + (5 a + 
    2 b) f^{(3)} r)))) 
\nonumber\\
&+& 8 \dot f r (-3 (18 a + 17 b) \ddot\theta \dot f \varepsilon r^2 + 
    \dot\theta (2 \varepsilon n (b - a (4 + 7 n)) + 3 r^2 + 
       3 \varepsilon r (-4 (2 a + b) \ddot f r 
\nonumber\\
&+&  \dot f (-10 a (2 + n) - 3 b (6 + n) + (10 a + 7 b) \dot f \dot\theta^2 r^3))))
\left.\right]\ .
\end{eqnarray}
\end{widetext}

\end{document}